\documentclass[12pt]{JHEP3}
\input epsf.tex
\epsfclipon

\usepackage{epsfig}
\usepackage{epstopdf}
\usepackage{epsf}
\input{epsf.sty}
\usepackage{graphicx,amsmath,amssymb}
\usepackage{epsfig,multicol}
\usepackage{bbold}
\allowdisplaybreaks

\usepackage{bbm,bm,amsmath,amssymb}

\def\bg{\begin{eqnarray}}
\def\nd{\end{eqnarray}}
\def\sin{{\rm sin}}
\def\cos{{\rm cos}}

\def\log{{\rm log}}

\newcommand{\pdfannotlink}{\pdfstartlink}
\newcommand{\be}{\begin{equation}}
\newcommand{\ee}{\end{equation}}
\newcommand{\ba}{\begin{eqnarray}}
\newcommand{\ea}{\end{eqnarray}}

\newcommand{\e}{\textrm{e}}

\newcommand{\w}{\wedge}
\renewcommand{\a}{\alpha}

\newcommand{\N}{\mathcal{N}}

\newcommand{\D}{\overline{\mathrm{D}3}}
\newcommand{\K}{K\"ahler }

\title{Renormalization Group Flow, Stability, and Bulk Viscosity in a Large $N$ Thermal QCD Model}
\author{Keshav Dasgupta, Maxim Emelin, Charles Gale, Michael Richard\\
\vskip.03in
Department of Physics, McGill University \\
3600 rue University, Montr\'{e}al, Qu\'{e}bec, Canada H3A 2T8\\
{\tt dasgupta.keshav8@gmail.com, maxim.emelin@mail.mcgill.ca}
~~{\tt gale@physics.mcgill.ca, michael.richard@mail.mcgill.ca}}

\date{\today}
\maketitle

\abstract{The ultraviolet completion of a large $N$ QCD model requires introducing new degrees of freedom at certain scale so that the UV behavior may become asymptotically 
conformal with no 
Landau poles and no UV divergences of Wilson loops. These UV degrees of freedom are represented by certain anti-branes arranged on the blown-up sphere of a warped resolved 
conifold in a way that they are separated from the other set of branes that control the IR behavior of the theory. This separation of the branes and the anti-branes creates 
instability in the theory. Further complications arise from the curvature of the ambient space. We show that, despite these analytical hurdles, stability may still be achieved by switching 
on appropriate world-volume fluxes on the branes. The UV degrees of freedom, on the other hand, modify the RG flow in the model. We discuss this in details by evaluating the flow from 
IR confining to UV conformal. Finally we lay down a calculational scheme to study bulk viscosity which, in turn, would signal the inherent non-conformality in this model.}

\begin{document}

\section{Introduction and a Brief Review of the Model \label{review}}
The theory of the nuclear strong interaction, Quantum Chromodynamics (QCD), becomes strongly  coupled as the energy scale is lowered. This fact renders practical calculations especially challenging, as the convergence of a perturbative expansion is no longer guaranteed. Traditionally, non-perturbative phenomena involving QCD could only be addressed through a small set of strategies. Some of these are: numerical techniques germane to a discretized version of the theory on a space-time lattice: lattice QCD \cite{Rothe:1992nt}, and  using the Operator Product Expansion to derive sum rules that circumvent some of the limitations of perturbation approaches \cite{Shifman:1978bx}.  In addition, several generations of experimental measurements have fuelled decades of phenomenological model-building. 

On the more formal side, two of the elements that have contributed to what has become a revolution in the theory of the nuclear strong interaction, is the observation that a non-Abelian gauge theory in the large number of colors limit has a perturbative expansion that matches that of a closed string theory \cite{'tHooft:1973jz}, and the celebrated AdS/CFT correspondence \cite{Maldacena:1997re}.  Even though an exact gravity dual to QCD -- with a finite number of colors -- is not at hand, theoretical constructions do exist that share some of its features, such as the appearance of a renormalisation scale. Those approaches offer the tantalizing prospect of being able to performed calculations in strongly-coupled QCD {\it analytically}. Many applications were concerned up to now with the many-body physics of the strongly-coupled quark-gluon plasma (QGP) \cite{CasalderreySolana:2011us} 
Indeed, experimental measurements performed at the Relativistic Heavy-Ion Collider (RHIC) at Brookhaven National Laboratory, and at the Large Hadron Collider (LHC) at CERN, have confirmed that the QGP is strongly coupled, in that it could very successfully be modelled using relativistic fluid mechanics \cite{Gale:2013da}. Much of the original excitement in the community  owed to the fact that the value of the effective shear viscosity over entropy density consistent with heavy ion data seemed to be approximately that inherent to a class of conformal field theories with
$\eta/s = 1/4\pi$ \cite{Kovtun:2004de}. It is also know that QCD is only approximately conformal, and hence has a non-vanishing coefficient of bulk viscosity \cite{Kapusta:2006pm}. It has recently also become evident that a finite bulk viscosity is also demanded by heavy-ion data \cite{Ryu:2015vwa}. 
The estimates for the precise value of the QCD transport coefficients are constantly being refined \cite{Gale:2013da}, but the hope of using string-based models in the study of hot and dense QCD remains. 

Having specified a context, detailed applications are not covered by this work but a model which can be used for such investigations is presented. Thus, from hereon we concentrate on one class of 
top-down supergravity models (for bottom-up approaches see for eg. \cite{kiritsis}). To study the supergravity dual of a model with renormalization group flow, one of the key question is how we can maintain {\it strong} 't Hooft coupling from UV to IR. This forms the 
basis of the construction of the Klebanov-Strassler model \cite{KS} where, at any given energy scale, there are an infinite number of gauge theory descriptions available out of which 
one (or a small set) is infinitely strongly coupled. The gravity dual from small $r$ to large $r$ corresponds to the set of these infinitely strongly coupled gauge theories from UV to IR, 
so that
at far IR it is the confining gauge theory whose dynamics is captured by the gravity dual.    

The Klebanov-Strassler model \cite{KS} provides a good description for the IR of the gauge theory. However there are UV issues related to Landau poles and divergences of the 
Wilson loops \cite{UVp} that require us to seek a UV completion of the Klebanov-Strassler model. The UV completion should be asymptotically conformal in terms of the 't Hooft 
coupling $\lambda$, so that it is asymptotically {\it free} in terms of the YM coupling $g^2_{YM}$. This way one of the requirements for a large $N$ QCD model may be 
easily taken care of. 

In \cite{fep} we managed to construct the gravity dual for a UV complete model of a large $N$ QCD. In fact we showed how thermal behavior could also be studied in this set-up. The 
detailed supergravity analysis may be succinctly presented in terms of three regions \cite{mia2}. Region 1 corresponds to the gravity dual of the IR regime of thermal QCD where 
confinement and deconfinement dynamics may be studied, whereas Region 3  corresponds to the asymptotically AdS region that captures the UV behavior of the theory. The intermediate 
region, Region 2, captures the dynamics of the theory when it is transforming from its deconfined stage to asymptotically conformal. In the gauge theory side the full UV 
completion requires us to insert $M$ anti-D5 branes in a set-up with $N$ D3 and $M$ D5 branes located at the south pole of a resolved sphere \cite{fep, mia2, uvcomplet}. The 
anti-D5 branes are separated from the D3/D5 branes and distributed on the upper half of the resolved sphere (see figure 1 in \cite{mesonic}). A natural question is then of the 
stability of the system against annihilation. The focus of section \ref{stability} is to show how we can stabilize the system using worldvolume fluxes (see also \cite{aalok} for a 
discussion on thermodynamic stablity of system, among other things).    

Once the system is stabilized, the gravity dual will have anti-D5 branes in Region 2 (the D3 and the D5 branes have transformed into metric and fluxes) alongwith a 
black hole\footnote{The complete backreacted geometry is, to our knowledge, first analysed in \cite{fangmia}. Later details appear in \cite{chempot}.}. 
This is the set-up where 
many of the thermal QCD calculations may now be performed. For example melting of the quarkonium states \cite{melting, patra}, QGP dynamics \cite{aalok}, effect of a chemical 
potential \cite{chempot}, energy loss of a moving quark \cite{fep}, transport coefficients including viscosity and entropy of the system \cite{fep, aalok, ramalata}, vector and 
scalar mesonic states \cite{mesonic} and renormalization group flow, among other things\footnote{Most of the coputations are performed in type IIB set-up. However one may also go to the 
mirror type IIA side to analyze the dynamics. See \cite{aalok} for details on this.}. 
The latter is discussed briefly in \cite{chempot} and \cite{uvcomplet} and in section
\ref{Rogi} we present a more detailed study. However one issue that has not been studied in the UV complete framework is the bulk viscosity $\zeta$. In section \ref{bulky} we lay out our 
computational scheme to study bulk viscosity in a UV complete model. We show that our model reproduces an umambiguous value of bulk viscosity, including the ratio of the bulk viscosity 
to the entropy density i.e $\zeta/s$. Our result is expressed in terms of a function that depends on the details of the UV completion of the model. In fact this function also governs the 
behavior of the bulk viscosity (and equivalently the ratio $\zeta/s$) for a different choice of the Schwinger-Keldysh quadrant used for the computation. Further details on the computation will be presented in  
\cite{richard}. We end with some discussions on the future prospects.   
 
\section{Stability, $\kappa$-Symmetry and Supersymmetry \label{stability}}

As discussed in section \ref{review} above, 
the UV completeness feature of the model requires anti-D5-branes, which may lead to instabilities due to D5-anti-D5 interactions and their eventual annihilation. The question of stabilizing brane-anti-brane configurations has been well explored on a flat background, but direct generalization to arbitrary curved spacetime is difficult. Still, the present setup is sufficiently simple that useful statements can be made regarding the stability of the model.

\subsection{First Look with Abelian Sources}

We will use the approach of \cite{0110039}, applied to the model in question. The goal is to check that a configuration of D5 and anti-D5 branes wrapped on a 2-sphere can be made stable by studying the $\kappa$-symmetry conditions of the branes. A crucial fact for this purpose is that worldvolume $\kappa$-symmetry on a supersymmetric background imply worldvolume supersymmetry. Our goal is to find fluxes on the brane and the anti-brane such that they preserve a common set of worldvolume supersymmetries, rendering the entire configuration BPS and therefore stable. This approach works for {\it probe} branes, for which we ignore the backreaction on the geometry.

The condition for a Dp-brane or anti-Dp-brane to be $\kappa$-symmetric is that there is a spinor satisfying:
\bg
\Gamma \epsilon = \pm \epsilon, 
\nd
with the $\pm$ used for the brane or anti-brane respectively and $\Gamma$ made up of contractions of the Levi-Civita tensor with all possible combinations of worldvolume gamma matrices and flux components. In type IIB it is given by:
\bg
&& \Gamma = \frac{\sqrt{|g|}}{\sqrt{|g+\mathcal{F}|}}\sum^\infty_{n=0}\frac{1}{2^n n!}\gamma^{j_1 k_1...j_n k_n}\mathcal{F}_{j_1 k_1}...\mathcal{F}_{j_n k_n} J^{(n)}_{(p)} \\ \nonumber
&& J^{(n)}_{(p)} = (-1)^n(\sigma_3)^{n+\frac{p-3}{2}}i \sigma_2 \otimes \frac{1}{(p+1)!\sqrt{|g|}}\epsilon^{i_1...i_{(p+1)}}\gamma_{i_1...i_{(p+1)}}.
\nd
All indices are worldvolume indices and the Pauli matrices in the second expression rotate the two same-chirality type IIB spinors into each other. $\mathcal{F}$ is the sum of the worldvolume gauge flux and the pullback of the background NS-NS $B$-field.

We now consider a D5 or anti D5 brane on a ${\bf R}^{3,1}\times {\bf S}_1^2 \times {\bf M}$, extended in the Minkowski directions and wrapping the ${\bf S}_1^2$, 
parametrized by $(\theta_1,\phi_1)$. The internal four-dimensional base ${\bf M}$ 
is locally of the form ${\bf T}_2^2 \times {\bf S}^1 \times {\bf R}_+$ 
to preserve Gauss' law. The global topology of ${\bf M}$ should be of the form ${\bf S}^3 \times {\bf R}_+$, but then we should worry about the curvature in the orthogonal directions 
to the five-branes. However we expect the result to be insensitive to the orthogonal metric, much like the supertubes construction which is stable regardless of the transverse 
metric \cite{supertube}. The local picture simplifies matter in the same way as in \cite{BBDK}, and the global extension doesn't change the story too much \cite{global}.  
For the present case, globally the two-torus ${\bf T}_2^2$ will become a sphere ${\bf S}^2_2$ and can be parametrized by coordinates ($\theta_2, \phi_2$) such that
the D5 brane and the anti-D5 brane are located at some fixed positions on the sphere. The metric on ${\bf M}$ can then be defined accordingly. On the other hand,     
we will assume the metric on the sphere ${\bf S}^2_1$ is diagonal but otherwise unspecified, with the full background metric taking the following form:
\bg\label{ninkay}
ds^2 = \frac{1}{\sqrt{h(r)}} ds^2_{0123} + \sqrt{h(r)} \left[dr^2 + f(r)^2 d\theta_1^2 + f(r)^2 \sin^2\theta_1~d\phi_1^2 + ds^2_{\bf M}\right], \nd
which is basically a simplified version of eq (2.11) in \cite{mia2} with $h(r)$ and $f(r)$ being two warp-factors whose precise functional forms will not be relevant for us. Note that 
\eqref{ninkay} should not be confused with the gravity dual, as it is the gauge theory side of the story\footnote{Topologically it is a resolved cone with the branes
distributed on the resolved sphere in a way described above. The gravity dual will be a resolved warped-deformed conifold with no branes other than the anti-D5 branes.}.  
The sphere parametrized by ($\theta_1, \phi_1$) on which we have the wrapped five-branes should shrink to zero size at $r = 0$, but we will consider an $f(r)$  that gives a finite size of the wrapped sphere. This is useful because, in the limit of vanishing size of the wrapped sphere, the fluxes on the sphere should become infinite to respect quantization rules \cite{DM2}.   
We want to introduce finite worldvolume fluxes $\mathcal{F}$ and $\bar{\mathcal{F}}$, satisfying the quantization conditions, 
on the brane such that their $\kappa$-symmetry equations are solved by the same spinor. As we will see, the dependence on $f(r)$ will drop out and the zero size limit can then be taken. 

Following a similar procedure to \cite{0110039} we turn on the $\mathcal{F}_{0\phi_1}$ and $\mathcal{F}_{\theta_1 \phi_1}$ flux components, which we will call $E$ and $B$ respectively. 
In this case the $\kappa$-symmetry condition becomes:
\bg \label{ksym}
\left(\gamma_{012345}\otimes \sigma_3 i \sigma_2 - E \gamma_{1234}\otimes i\sigma_2 - B\gamma_{0123}\otimes i \sigma_2 \over \sqrt{\left|g + \frac{1}{h^2} B^2  + \frac{f^2}{h}E^2\right|}\right)\epsilon 
= \pm \epsilon.
\nd
One simple way to solve this is to cancel the first two terms in the numerator against each other. This requires:
\bg \label{d1}
\gamma_{05}\otimes i\sigma_3  \sigma_2 \epsilon = iE  \sigma_2 \epsilon.
\nd
We need to choose the electric field $E$ such that this expression has solutions. 
Note that $\gamma_i$ are \emph{worldvolume} gamma matrices, related to usual flat space gamma matrices $\Gamma_i$ by the worldvolume vielbeins. Choosing $E=\sqrt{-g_{00} g_{\phi_1 \phi_1}}$, this 
becomes:
\bg
\Gamma_{05}\otimes\sigma_3  \epsilon = -\epsilon.
\nd
Once this condition is satisfied, the original expression \eqref{ksym} for the 5-brane becomes:
\bg \label{d3}
h \gamma_{0123}\otimes i \sigma_2 \epsilon&= \Gamma_{0123}\otimes i \sigma_2 \epsilon=\mp \frac{\left|B\right|}{B}\epsilon.
\nd
Since $g_{\theta_1 \phi_1}=0$ from \eqref{ninkay}, and choosing opposite values of $B$ for the brane and the anti-brane, 
this becomes exactly the expression for a D3 brane stretched along the Minkowski directions and positioned at any point on the two-sphere ${\bf S}^2_2$. 
It's not difficult to see, using properties of gamma matrices, that \eqref{d1} and \eqref{d3} can be solved simultaneously. 

Unfortunately it is easier said than done, as there is a glaring problem with this result. The required total flux: 
\bg \label{flux}
\mathcal{F}=e_0 \wedge e_{\phi_1} + \frac{B}{\sqrt{h} ~f^2~ \sin\theta_1} ~ e_{\theta_1} \wedge e_{\phi_1}, 
\nd 
is not closed, so it can't be the field strength of the worldvolume gauge potential, nor can it come from a background NS-NS $B$-field, since the resulting 3-form flux doesn't satisfy 
$d \star H_3 =0$, and would require spacelike sources.
This means the configuration of a D5-brane and an anti-D5 brane on the two-sphere ${\bf S}^2_2$ cannot be stabilized in the usual way by abelian 
fluxes\footnote{The total flux equation \eqref{flux} implies that $dF \ne 0$, so another way to interpret this would be to take 
magnetic sources into account. These magnetic sources cannot be point-like, compared to what we have in four space-time dimensions. 
Assuming $\ast F = dC_3 + ....$, we can modify the world volume action by including sources as: 
$$ S = {1\over g^2_{YM}} \int_{{\bf \Sigma}_6} F \wedge *F + \int_{{\bf \Sigma}_6}  C_3 \wedge \delta^3(\vec{\bf x}).$$ 
Thus in the presence of these sources, one might make sense of \eqref{flux} albeit in a bit contrived way. Simpler analysis 
exists, as we elucidate in the following section.}.   
However once we replace ${\bf S}^2_1$ by a torus ${\bf T}_1^2$, as in \cite{DHHK}, abelian fluxes do stabilize a brane-anti-brane system.

\subsection{Non-Abelian Sources and $\kappa$-Symmetry}

The solution to this conundrum can come from the fact that our setup contains multiple branes and the corresponding worldvolume action is non-abelian. Indeed, in a non-abelian gauge 
theory the field strength is given by:
\bg
F= dA+A\wedge A,
\nd
and the second term is not closed. However, in moving to the non-abelian case we have to deal with additional complications. All the quantities in the expressions we derived so far now carry adjoint $SU(M)$ gauge indices and the form of the $\kappa$-symmetry matrix will also change. Therefore even though we can construct a worldvolume flux of the form \eqref{flux}, it's no longer obvious that this flux is the one required to restore supersymmetry to the system or what its gauge components should be! The situation is complicated by the fact that the full non-abelian $\kappa$-symmetry transformation, like the non-abelian DBI action, is not known. We will, however present some hints that a non-abelian gauge flux may be capable of restoring supersymmetry to the system. While the full $\Gamma$ matrix in the $\kappa$-symmetry transformation is not known, one can compute it order by order in the worldvolume flux following \cite{0011018v1}. 
The result to second order is:
\bg \label{nonabgamma}
\Gamma^{AB} &=& \Gamma^{(0)} \Bigg\{ \sigma_1 \delta^{AB} + i \sigma_2 d^{ABC} \frac{1}{2}\gamma^{kl} F^C_{kl} 
- \sigma_1 \mathcal{A}^{ABCD} \frac{1}{2} \gamma_{ij}F^{ik C}F_{k}^{\,\,jD} \nonumber\\  
&& ~~~~~~~~~-\sigma_1 \mathcal{S}^{ABCD}\left( \frac{1}{8} \gamma_{ijkl}F^{ij C} F^{kl D} - \frac{1}{4} F^C_{kl}F^{kl D}\right)\Bigg\}, 
\nd
where ${\cal S}^{ABCD}$ and ${\cal A}^{ABCD}$ are defined using the $SU(M)$ Lie algebra structure constants $d^{ABC}$ etc, in the following way:
\bg \label{altyler}
\mathcal{S}^{ABCD} =  d^{AE(C}d^{D)EB}, ~~~~~~~~ \mathcal{A}^{ABCD} = d^{AE[C}d^{D]EB}.
\nd
A few things need to be pointed out. First, since our field strengths are of order one, simply achieving $\kappa$-symmetry at any finite order is not at all sufficient to conclusively declare the system to be stable. The purpose of this calculation is to look for hints that it is possible, but a full proof would require the full non-abelian action.

Second, we will no longer have the automatic normalization of the $B$-field. This is due to the lack of the determinant factor in front of our $\kappa$-symmetry matrix. We can either set 
${\rm Tr}~ B^2=\pm 1$, further invalidating the order by order approach, or hope that the full non-abelian $\kappa$-symmetry transformation has this normalization, but it gets hidden 
in the order-by-order expansion. We will see hints that this may be the case from the term at second order in \eqref{nonabgamma}, namely:
\bg\label{lison}
{1\over 4} \sigma_1 \mathcal{S}^{ABCD} F^C_{kl}F^{kl D}, \nd
but as pointed out in \cite{0011018v1}, there's no obvious factorization that takes place. Indeed, if we knew the exact factorization of the $\Gamma$ matrix, it would amount to knowing 
the non-abelian DBI action.

Before proceeding, it is worth taking the time to set up some notation. The generators $t^A$, with $1\leq A \leq M^2-1$, of $SU(M)$ expressed in the fundamental representation can be split into diagonal, off-diagonal symmetric and anti-symmetric matrices. We can label these subsets of generators by $t^{(d)}, t^{(s)}, t^{(a)}$ respectively and will order our basis accordingly, so that $t^1, ... ,t^{M-1}$ are diagonal, $t^M, ... , t^{(M^2+M-2) /2}$ are symmetric and the rest are anti-symmetric. 

We will also pick a particular (quite standard) basis for the non-diagonal generators, such that each basis element has only one non-zero entry in the upper triangle (and the corresponding entry in the lower triangle). We then label the symmetric generator with the non-zero entry in the $i$-th row, $j$-th column as $t^{(s)}_{ij}$ and similarly for the anti-symmetric generators. The order for this basis will be given by:
\bg
\bigg\{t^M, t^{M+1}, ... , t^{2M-1}, t^{2M}, ... \bigg\} = \bigg\{t^{(s)}_{12}, t^{(s)}_{13} , ... , t^{(s)}_{23}, t^{(s)}_{24}, ...  \bigg\},
\nd
and similarly for the $t^{(a)}$'s. Additionally, the non-vanishing symmetric and anti-symmetric structure constants, defined by:
\bg
t^A t^B= \left(d^{ABC}+i f^{ABC}\right) t^C, \nd
can be deduced from the symmetry properties of (anti-)commutators. They are of one of the following forms:
\bg
&f^{(dsa)}, \; f^{(ssa)},\; f^{(aaa)} \nonumber \\
&d^{(ddd)},\; d^{(sss)},\; d^{(dss)},\; d^{(saa)},\; d^{(daa)},
\nd
and all others related by permutation.

It will be convenient to think of these structure constants as a collection of matrices labeled by their last index and acting on the adjoint representation. We will define 
$F^{(C)}$ and $D^{(C)}$ such that:
\bg
\left[F^{(C)}\right]^{AB} = f^{ABC}, ~~~~~~~ \left[D^{(C)}\right]^{AB} = d^{ABC}, 
\nd
the $\mathcal{S}$ and $\mathcal{A}$ tensors can be expressed as commutators and anti-commutators of these matrices.

We now return to our problem. For simplicity we switch on gauge field components $A_0$ and $A_{\phi_1}$ such that they are functions of $\theta_1$ coordinate only. 
With adjoint indices written explicitly, our ansatz for the flux becomes:
\bg \label{fluxcomp}
F^A =B^A e^\theta \wedge e^\phi +E^A e^0 \wedge e^\phi 
=\partial_\theta A_{\phi}^A e^\theta \wedge e^\phi + f^{ABC}A_0^B A_\phi^C e^0 \wedge e^\phi,
\nd
from which it is obvious that the two spacetime components of the flux must also be different generators of $SU(M)$. Defining:
\bg
\mathcal{E} = \sum_A E^A D^{(A)}, ~~~~ \mathcal{B} = \sum_A B^A D^{(A)},
\nd
which are now matrices with two adjoint indices, the $\kappa$-symmetry condition in this notation and using our ansatz \eqref{fluxcomp} becomes: 
\bg \label{kappaNA}
&&\bigg(\sigma_1 \otimes \gamma_{012345} \otimes \mathbb{1} + \, i \sigma_2 \otimes\gamma_{1234} \otimes \mathcal{E} + \,  i \sigma_2  \otimes \gamma_{0123} \otimes  \mathcal{B}\nonumber \\ 
&& ~~~- \, \frac{1}{2} \sigma_1 \otimes \gamma_{012345} \otimes (\mathcal{E}^2 + \mathcal{B}^2) + \, \sigma_1 \otimes \gamma_{1235} \otimes [\mathcal{E}, \mathcal{B}] \bigg)\epsilon = 
\pm \epsilon,
\nd
where the spinor now carries an adjoint index, acted on by the last factor in the tensor products, as well as the usual $SU(2)$ multiplet and spinor indices. The other second-order term involving $\mathcal{S}$ vanishes for our ansatz by anti-symmetry in the lorentz indices. We can also force the $\mathcal{A}$ term to vanish by choosing $E^A$ and $B^A$ such that $\mathcal{E}$ and $\mathcal{B}$ commute, at least on a subspace of the adjoint representation. The spinors satisfying this condition will have to then lie in that subspace. 

This can be achieved, for example, by choosing $E=t^{M}=t^{(s)}_{12}$ and $B=B t^{(M^2+M)/2}=B t^{(a)}_{12}$, i.e. two of the generators of an $SU(2)$ subgroup acting on the first two components of the fundamental representation. The non-vanishing anti-commutators involving these generators are:
\bg\label{d}
&&\{t^{(s)}_{12}, t^{(s)}_{12} \} \in t^{(d)} \nonumber  \\
&&\{t^{(a)}_{12}, t^{(a)}_{12} \} \in t^{(d)} \nonumber  \\
&&\{t^{(s)}_{12}, t^{(s)}_{1j} \} \propto t^{(s)}_{2j} \nonumber \\
&&\{t^{(s)}_{12}, t^{(s)}_{2j} \} \propto t^{(s)}_{1j}  \nonumber\\
&&\{t^{(s)}_{12}, t^{(a)}_{1j} \} \propto t^{(a)}_{2j}\nonumber  \\
&&\{t^{(s)}_{12}, t^{(a)}_{2j} \} \propto t^{(a)}_{1j} \nonumber  \\
&&\{t^{(a)}_{12}, t^{(a)}_{1j} \} \propto t^{(s)}_{2j}  \nonumber  \\
&&\{t^{(a)}_{12}, t^{(a)}_{2j} \} \propto t^{(s)}_{1j}, 
\nd
where the actual numerical coefficients are the same between the expressions involving $t^{(s)}_{12}$ and the analogous $t^{(a)}_{12}$ expressions, 
meaning that $\mathcal{E}$ and $\mathcal{B}$ have the following schematic form:
\bg \label{matrices}
   \mathcal{E} \propto
 {\begin{bmatrix}  
  0&  D & 0 & 0 & 0 \\   D^{{}^\intercal} & 0 & 0 & 0 & 0  \\  0 & 0 & C & 0 &  0  \\ 0&0&0&0&0 \\  0 & 0 & 0  & 0 &  C    \end{bmatrix} },  
\qquad \qquad        \mathcal{B} \propto
 {\begin{bmatrix}
  0&  0 & 0 & D & 0 \\   0 & 0 & 0 & 0 & 0  \\  0 & 0 & 0 & 0 &  C  \\ D^{{}^\intercal} &0&0&0&0 \\  0 & 0 & C  & 0 & 0     \end{bmatrix} } 
\nd
where $D$ is a $1\times (n-1)$ block containing the constants coming from the first two equations in \eqref{d}  and $C$ is a symmetric square block of size $(M^2 + M - 2)/2$ containing the constants coming from the remaining equations in \eqref{d}. It's not too difficult to see that these matrices commute when acting on almost the whole adjoint representation except the subspace spanned by $t^{(s)}_{12}, t^{(a)}_{12}$, so the anti-symmetric term in the $\Gamma$ matrix drops out if we restrict $\epsilon$ to be orthogonal to that subspace. Furthermore, since all the $SU(2)$ subgroups of $SU(M)$ are related by a change of basis, we can pick two generators of any $SU(2)$ subgroup to be our field strengths.
This leaves us with:
\bg \label{kappasimple}
\bigg[\sigma_1 \otimes \gamma_{012345} \otimes \left(\mathbb{1} - {\mathcal{E}^2 + \mathcal{B}^2 \over 2}\right) 
+ \, i \sigma_2 \otimes\gamma_{1234} \otimes \mathcal{E} + \,  i \sigma_2  \otimes \gamma_{0123} \otimes  \mathcal{B} \bigg]\epsilon = \pm \epsilon. \nonumber\\ 
\nd
Interestingly, 
this looks exactly like the expansion of \eqref{ksym} to second order with $E, B$ becoming the matrices $\mathcal{E}$ and $\mathcal{B}$.  If we assume that the quadratic terms in the field strengths come from a similar expansion, we only need to satisfy the condition to linear order in the field strengths. The procedure is then analogous to the abelian case, resulting in the following two conditions:
\bg \label{NAD12}
\sigma_3\otimes \Gamma_{05} \otimes \mathcal{E} \, \epsilon = \epsilon, ~~~~~
i\sigma_2\otimes\Gamma_{0123}\otimes \mathcal{B} \, \epsilon = \mp \epsilon,
\nd
which are now satisfiable for field strengths of the form \eqref{fluxcomp}. The spinor $\epsilon$ that will satisfy these conditions has to be an eigenvector 
of $\mathcal{E}$ and $\mathcal{B}$, i.e. be made up of appropriately placed blocks that are eigenvectors of the $C$ blocks in \eqref{matrices}. Note that this spinor will indeed be orthogonal to the subspace on which $\mathcal{A}$ doesn't vanish. This way the stability that we seek in 
this configuration may be achieved\footnote{It may be interesting to note that the anti-brane issues pointed out in \cite{bena}, in the gravity dual of our framework,  
are expected to be absent because our system is supersymmetric 
and therefore stable.}. 

\section{Renormalization Group Flow \label{Rogi}}

In the previous section we discussed how a stable configuration of D5 and anti-D5 branes may be constructed using world-volume fluxes. As emphasized above, such a configuration is at least
necessary to have a UV completion of the Klebanov-Strassler model. The UV complete model also has fundamental matter (for example ``quarks''), which are generated by inserting $N_f$
D7-branes to the gravity dual of system i.e in Region 3. 
However to avoid the resulting Landau poles, the {\it full} UV completion requires us to add anti-D7 branes to the system \cite{mia2}. This can again be 
stabilized by world-volume fluxes, although their effects are $g_sN_f$ suppressed. The anti-D7 branes remove all the log $r$ pieces that lead to landau poles and keep only the $r^{-n}$ 
terms so that there are no UV issues in the field theory side. 
    
Therefore to summarize, the model presented in above is a modification of the Klebanov-Strassler (KS) geometry with the addition of seven-branes to include fundamental flavors 
and an asymptotically AdS cap, which brings the corresponding field theory to a UV fixed point. The far UV is then governed by a $SU(N+M) \times SU(N+M)$ gauge theory, with a 
{\it walking} RG flow governed mainly by the distribution of the flavors (i.e the seven-branes in the dual side). At certain scale, the gauge group gets Higgsed to $SU(N+M) \times SU(N)$    
whence the field theory undergoes a cascade of Seiberg dualities decreasing the number of colors as we flow to the IR until there is only a confining $SU(M)$ theory 
remaining\footnote{With $N$ and $M$ appropriately chosen. For example one choice is $N = (k-1)M$ such that the UV gauge group is $SU(kM) \times SU(kM)$ which gets Higgsed to 
$SU(kM) \times SU((k-1)M)$ at some intermediate scale giving us the minimally supersymmetric $SU(M)$ gauge theory at far IR.}.   
In the following, section \ref{sedual}, 
we review some features of the cascade and clarify which features of the renormalization group flow we can expect to see from the gravity side. 
For a more detailed review see \cite{0505153}. We then add the fundamental flavors in section \ref{fundu}, analyze the B-fields and dilaton in section \ref{dilatone}, followed by the 
study of the full RG flow from UV to IR in section \ref{rgfull}.

\subsection{Review of Seiberg Duality in KS Model \label{sedual}}

The Klebanov-Strassler field theory is an $\mathcal{N}=1$ gauge theory with gauge group $SU(N+M)\times SU(N)$. There are two bifundamental chiral multiplets, $A_1, A_2$ and two anti-bifundametal chiral multiplets $B_1, B_2$. They are coupled through a classical superpotential:
\bg
W = h \, \, {\rm tr}~ {\rm det}_{i,j} (A_i B_j), \qquad i,j\in{1,2}
\nd
For $M=0$ this is the Klebanov-Witten model \cite{KW}, which has a two-dimensional surface of fixed points. For $M\neq 0$ the beta functions for the couplings are:
\bg
&& \beta_\eta = \eta(1+2\gamma_0) \\
&& \beta_{g_1} = - {g_1^3\over 16 \pi^2}\left[\frac{3(N+M)-2N(1-\gamma_0)}{1-g_1^2 N/8\pi^2}\right], ~~~
\beta_{g_2} = - {g_1^3\over 16 \pi^2}\left[\frac{3N-2(N+M)(1-\gamma_0)}{1-g_2^2 N/8\pi^2}\right], \nonumber 
\nd
where $\eta$ is the dimensionless version of the quartic coupling $h$ and $\gamma_0$ is the anomalous dimension of the chiral multiplets, which is a generally an 
unknown function of all the couplings.

The gauge couplings flow is given by the NSVZ beta function, with indices in each factor of the gauge group acting as flavor indices for the other factor, so $g_1$ sees $2N$ flavors, while $g_2$ sees $2(N+M)$ flavors. There are no fixed points for which all three couplings are non-zero. For $g_1=0, \eta=0$ there is assumed to be a Seiberg fixed point at non-zero $g_2$ 
if $2M < N$. Likewise there's a Seiberg fixed point at non-zero $g_1$ for $g_2=0,\eta=0$. The former is stable in the $\eta$ direction, but unstable in the $g_1$ direction, while the latter is stable in the $g_2$ direction but unstable in the $\eta$ direction since $\gamma_0<-\frac{1}{2}$. In the $g_2=0$ plane the theory essentially becomes an $SU(N+M)$ gauge theory with $2N$ flavors of (anti-)fundamental chiral multiplets, $A_i$($B_i$) transforming in the (anti-)fundamental representation of the flavor group and a quartic coupling between them with a coupling constant $\eta$. 

At the Seiberg fixed point, this theory has a dual description via Seiberg duality \cite{9411149, 0505153}. The dual theory is an $SU(2N -(N+M))=SU(N-M)$ theory with $2N$ flavors, with its own ``dual'' chiral multiplets (call them $\widetilde{A}_i,\widetilde{B}_i$) transforming in the opposite representation of the flavor symmetry group compared to the original theory. This theory also contains gauge-neutral ``meson" fields ${\cal M}$ transforming in a bifundamental representation of the flavor group, which couple to the chiral multiplets via a 
$y {\cal M} \widetilde{A}\widetilde{B}$ superpotential. These ${\cal M}$ fields are dual to bilinear combinations $AB$ of the chiral multiplets of the original theory. 
At the Seiberg fixed point of the original theory, these fields are massless, but the quartic coupling in the original theory is dual to a relevant mass term for the 
${\cal M}$ fields. As we flow to the IR, $\eta$ grows, the now massive ${\cal M}$ fields get integrated out and the dual theory reduces to an $SU(N-M)$ theory with chiral multiplets 
$\widetilde{A}_i, \widetilde{B}_i$ at its own Seiberg fixed point.

\begin{figure}[t]
  \centering
    \includegraphics[width=0.8\textwidth]{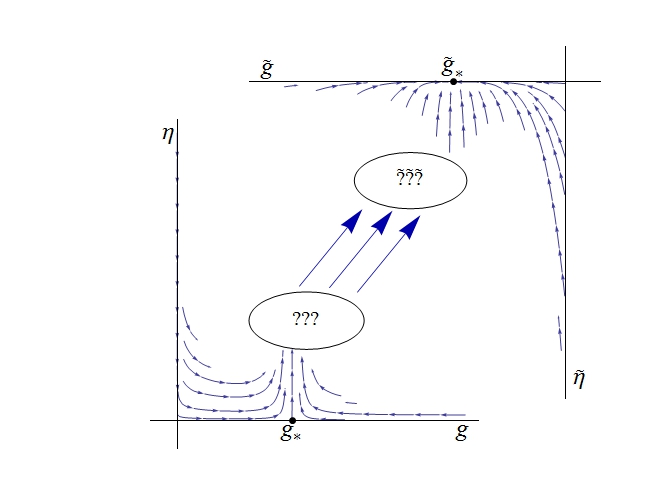}
  \caption{Qualitative features of the RG flow of the $SU(N+M)$ theory and the Seiberg dual $SU(N-M)$ theory. The ``???'' represent our ignorance of the behavior outside a neighborhood of the Seiberg fixed points.}
\label{seibdual}
\end{figure}

The original theory, on the other hand, \emph{gains}, a set of gauge-neutral bifundamental fields, which get ``integrated in''. Indeed, we can rewrite the quartic coupling of our theory as:
\bg
y \, \, {\rm tr} \left(A_1 B_1 \widetilde{M}_{22}+ A_2 B_2 \widetilde{M}_{11} - A_1 B_2 \widetilde{M}_{21}- A_2 B_1 \widetilde{M}_{12}\right) 
+ \frac{2y}{\sqrt{h}} \left(\widetilde{M} \widetilde{M}\right),
\nd
for some auxiliary field $\widetilde{M}$, where we suppressed both the ``color'' $SU(N+M)$ indices, under which the $\widetilde{M}$ are neutral and the ``flavor'' $SU(N)$ indices under 
which $\widetilde{M}$ transform in the bi-fundamental representation. The last term has index contractions analogous to the first term.

The mass term for this new field field is actually \emph{irrelevant} (since $h$ is relevant), so treating the $\widetilde{M}$'s as very heavy dynamical fields, they will become massless in the IR. This is to be expected. Seiberg duality is \emph{exact} at the Seiberg fixed point of the original theory. As we then flow further to the IR toward the Seiberg fixed point of the dual theory with ${\cal M}$ integrated out, Seiberg duality continues to hold \emph{exactly} along that line. Since Seiberg duality is a duality between a theory \emph{without} ``meson" fields and a theory \emph{with} ``meson" fields, if one description loses them, the other must gain them.

Finally, we note that from a field theory perspective, the behavior of the gauge coupling in either description is generally unknown for non-zero quartic coupling outside the neighborhood of their respective Seiberg fixed point as shown in {\bf fig \ref{seibdual}}. 
Similarly, there is no simple relationship between the gauge coupling and its dual. In particular, the dual gauge coupling isn't even guaranteed to be finite at the original fixed point. It's possible that somewhere along the flow between the two fixed points, one description's gauge coupling diverges and is in a confined phase by the time we reach the fixed point of its dual. We will indeed see such divergences in our gravity analysis.

Once the dual description reaches its Seiberg fixed point, the full theory is $SU(N-M)\times SU(N)$. With gauge couplings $\widetilde{g}_1, g_2$ 
(the second coupling constant remained unchanged). This Seiberg fixed point is unstable in the $g_2$ direction and the flow takes us to the $SU(N)$ Seiberg fixed point 
at $\widetilde{g}_1 = 0, g_2\neq 0$, which is again unstable towards developing a quartic coupling, growing a set of massless ``meson'' fields and forcing us to Seiberg dualize to an 
$SU(N-M)\times SU(N-2M)$ theory etc. This process repeats until we wind up, through a judicious choice of the number of colors in the UV, with an $SU(M)\times SU(0) = SU(M)$ theory which ultimately confines without offering us a Seiberg dual theory to transform to.

Obviously, to go down the entire cascade we should avoid hitting the Seiberg fixed points exactly, which is not that difficult, due to their instability. One can talk about ``weakly coupled'' RG flows, which pass very close to the fixed points, so the gauge couplings become small at least occasionally. In this regime the flow lingers near the fixed points over large energy ranges and then quickly flows toward the next fixed point forcing a change of variables via Seiberg duality. There are however also ``strongly coupled'' flows, which miss the fixed points by a large margin, constantly have large coupling constants and therefore don't really have a useful description in terms of any of the $SU(N+M)\times SU(N)$ theories. It is in this regime that the gravity description of the theory becomes good.

\subsection{Effects of Fundamental Flavors and UV Completion \label{fundu}}

The two major differences between the model described in section \ref{stability} and the KS model are the presence of fundamental flavors from the seven-branes 
and a UV completion to the theory. As we discussed earlier, 
where rather than staying on the duality cascade at all energy scales we start instead with an $SU(N+M)\times SU(N+M)$ theory and Higgs one of the 
$SU(N+M)$'s at a suitable energy scale so as to land on a duality cascade that ends as a confining $SU(M)$ theory in the far IR. 
The addition of $N_f$ fundamental matter fields to the theory simply changes how many flavors in total each factor of the gauge group sees. This influences our choice of initial gauge group 
rank, since we still want to end up with a confining theory in the IR. Also, a sufficiently large $N_f$ can influence the last few steps of the cascade by forcing the gauge theory 
outside of its conformal window thus removing the Seiberg fixed points. The latter effect already happens even without the addition of extra flavors. 
For example by the time the flow reaches an $SU(3M)\times SU(2M)$ theory the $SU(3M)$ sees $4M$ flavors, so $N>\frac{2}{3}N_f$, which is outside the conformal window. 
For more details regarding these subtleties, see \cite{0505153}. Regardless, at strong coupling we are constantly far from the Seiberg fixed points, so these details will not be 
captured by the analysis of the gravity dual.

\begin{figure}[t]\centering \includegraphics[width=0.8\textwidth]{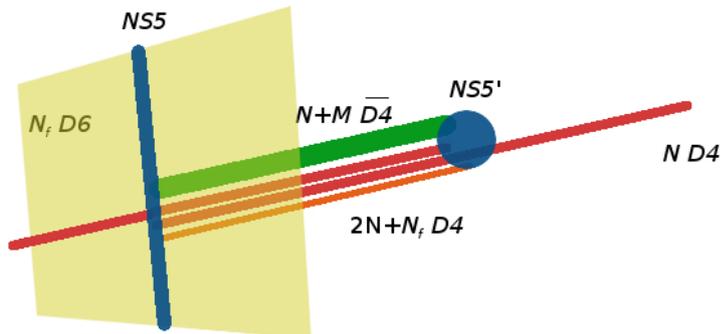}\vskip-2.0in \caption{The IR gauge group after a Seiberg duality in the presence of $N_f$ flavors in the type IIA dual picture. The flavors are D6-branes, divided into two halves by one of the NS5-brane. The other NS5'  brane crossing the D6-branes create $2N + N_f$ extra D4-branes. Together with the $N + M$ anti-D4 branes, the gauge group after Seiberg duality becomes $SU(N) \times SU(N - M + N_f)$ consistent with an actual gauge theory computation.}\label{IIAbranes}\end{figure}

A more careful analysis however reveals additional subtleties. On one hand from the gravity dual perspective, the flavor seven-branes are arranged in Region 3 in a 
way as to avoid creating Landau poles in the UV. As we discussed briefly earlier, this amounts to putting D7 and anti-D7 branes\footnote{The correct picture is to include both local and 
non-local seven-branes \cite{mia2}.} 
so that the background fields do not have any 
log $r$ behaviors. This means the seven-branes are arranged via Ouyang embedding \cite{ouyang} with, as discussed in section 2.3 of \cite{mia2}, {\it bound} states of 
D7 and anti-D7 branes in Region 3 and D7 branes in Regions 1 and 2. The backreactions of the D7-branes in Regions 1 and 2 now restrict the number of D7-branes to be less than 
24 \cite{vafa}. Thus $N_f$ cannot be too large for our case, and since $g_s << 1$, the effects of the seven-branes are $g_sN_f$ suppressed. 

On the other hand, the IR physics do change a bit from what we studied above. This is most succinctly presented in the T-dual type IIA language\footnote{This is the T-dual of the 
brane construction leading to the confining gauge theory i.e the T-dual of the wrapped D5-branes on the vanishing two-cycle of a conifold in the presence of the D7-branes. The T-dual of 
Region 1 in the gravity dual will lead to a somewhat different story that we will not elaborate here.} 
as shown in {\bf fig \ref{IIAbranes}}. 
For simplicity we will only discuss the physics in Region 1 i.e the cascade part of the story so as to avoid the complications that may arise from the anti-D5 branes.  

In the T-dual type IIA side, the flavor D7-branes become D6-branes in a configuration of intersecting NS5-branes oriented as in \cite{DM1, DM2} with $M$ D4-branes in between. The D6-branes
are divided into two halves by one of the NS5-brane, as shown in {\bf fig \ref{IIAbranes}}. Once we cross the NS5-branes, the $M$ D4-branes turn into $M$ anti-D4 branes, but we also 
get additional $N_f$ D4-branes from the Hanany-Witten \cite{hanwit} brane creation process from the $N_f$ D6-branes (see also \cite{ouyang}). This way after one Seiberg duality the 
gauge group changes to $SU(N) \times SU(N - M + N_f)$. 
At the end of the cascade, the far IR picture remains similar to what we expect from the brane construction: a ${\cal N} = 1$ supersymmetic Yang-Mills $SU(M)$ theory with $N_f$ fundamental 
flavors. This is exactly the story that also emerges from the gravity dual, which we elaborate next.

\subsection{Behavior of the NS B-field and the Dilaton \label{dilatone}}

In the gravity dual, the relevant quantities to analyze are the NS B-field and the dilaton. This is because
the gauge coupling constants are related to the dilaton and the B-field of the dual gravity description,  both of which have been computed in \cite{KW, DM1, fep}, by:
\bg \label{dileshwar}
\frac{4\pi^2}{g_1^2}+\frac{4\pi^2}{g_1^2}&=&e^{-\Phi} \nonumber\\
\frac{4\pi^2}{g_1^2}-\frac{4\pi^2}{g_1^2}&=&\frac{e^{-\Phi}}{2\pi}\left[\left(\int_{S^2} {\widetilde B}_2 \mod 2\pi\right)- \pi \right], 
\nd 
where $\Phi$ is the dilaton and ${\widetilde B}_2$ 
is the NS B-field threading Regions 1, 2 and 3. As discussed in eqn (2.75) of \cite{chempot}, the total field strength of the NS B-field, $H_3$, is 
a complicated three-form that can be expressed as:
\bg\label{shala}
H_3 & = & {\bf F}_1 \left(\sin~\theta_1~d\theta_1 \wedge d\phi_1 + {\bf A}_2~ \sin~\theta_2~d\theta_2 \wedge d\phi_2\right)\wedge dr  \nonumber\\
&& ~~+  \left({\bf F}_2 ~dr \wedge e_\psi + {\bf F}_3 ~de_\psi\right) \wedge \left({\rm cot}~\theta_1~d\theta_1 + {\bf A}_2 ~{\rm cot}~\theta_2~d\theta_2\right), \nd
where ($\theta_i, \phi_i, r, \psi$) are the coordinates of the resolved warped-deformed conifold, with ${\bf F}_i$ and ${\bf A}_i$ are functions of all the six coordinates as well as the 
resolution parameter $a^2$ (which is also defined in \cite{chempot}). 
Their precise functional
forms may be read up easily from eqn (2.75) in \cite{chempot}. Note that $H_3 = d{\widetilde B}_2$, 
and so it is a closed three-form. This closure implies certain conditions on all the parameters 
involved in \eqref{shala}, as may be inferred from eqns (2.78) and (2.79) of \cite{chempot}. 

What we now need are the precise forms of the NS B-field, ${\widetilde B}_2$, the dilaton ${\Phi}$ and the resolution parameter $a^2$. We will start with the NS B-field. It is given by
eqn (2.91) of \cite{chempot} that we reproduce here for convenience:
\bg\label{hityy}
{\widetilde B}_2 = B_2(r, \theta_i) + \left(g_s M\right) (g_sN_f) \left({g_s M^2\over N}\right)\Big[{\cal B}_2(r) + g_s N_f {\cal C}_2(r, \theta_1)\Big]. \nd
However in the limit given by eqn (2.38) of \cite{chempot}, the second and the third terms of \eqref{hityy} are suppressed by powers of $\epsilon$ (the small parameter which controls the relative scaling of $g_sM, g_sM^2/N$ and other small quantites in our limit) as in eqn (2.92) of \cite{chempot}. 
Similarly, one may show that the resolution parameter $a^2$ is given by the constant piece $a_0^2$ with the other parts suppressed as in eqn (2.92) of \cite{chempot}. This means
the NS-NS B-field takes the following form (see also eqn (2.88) of \cite{chempot}):
\bg\label{b2}
 B_2 &=& \left(b_1(r)\cot\frac{\theta_1}{2}\,d\theta_1+b_2(r)\cot\frac{\theta_2}{2}\,d\theta_2\right)\wedge e_\psi\\ \nonumber
  & + &\left[\frac{3g_s^2MN_f}{4\pi}\,\left(1+\log(r^2+9a_0^2)\right)\log\left(\sin\frac{\theta_1}{2}\sin\frac{\theta_2}{2}\right)
    +b_3(r)\right]\sin\theta_1\,d\theta_1\wedge d\phi_1\\ \nonumber 
  & - & \left[\frac{g_s^2MN_f}{12\pi r^2}\left(-36a_0^2+9r^2+16r^2\log r+r^2\log(r^2+9a_0^2)\right)
    \log\left(\sin\frac{\theta_1}{2}\sin\frac{\theta_2}{2}\right)+b_4(r)\right]\\ \nonumber
  & & \qquad\qquad \times ~\sin\theta_2\,d\theta_2\wedge d\phi_2,  
\nd
where we take the resolution parameter $a^2 \approx a_0^2$ to be approximately a constant, and express the coefficients $b_i$ as:
\bg \label{bcoeffs}
&& b_1(r) = \frac{g_S^2MN_f}{24\pi(r^2+6a_0^2)}\big(18a_0^2+(16r^2-72a_0^2)\log r+(r^2+9a_0^2)\log(r^2+9a_0^2)\big) \nonumber\\
&&  b_2(r) = -\frac{3g_s^2MN_f}{8\pi r^2}\big(r^2+9a_0^2\big)\log(r^2+9a_0^2)\nonumber\\
&& b_3(r) = \int_{a_0}^r dy \Bigg\{\frac{3g_sM y}{y^2+9a_0^2} + \frac{g_s^2MN_f}{8\pi y(y^2+9a_0^2)}\Big[-36a_0^2-18 a_0^2\log ~a_0^2 + 34 y^2\log~ y \nonumber\\
  & & \qquad\qquad\qquad\qquad\qquad\qquad+(10 y^2+81a_0^2)
    \log(y^2+9a_0^2)\Big]\Bigg\}\nonumber\\
&& b_4(r) = -\int_{a_0}^r dy \Bigg\{\frac{3g_sM(y^2+6a_0^2)}{\kappa y^3} + \frac{g_s^2MN_f}{8\pi\kappa y^3}\Big[18a_0^2-18(y^2+6a_0^2)\log ~a_0^2 \nonumber\\ 
  & & \qquad\qquad\qquad+(34 y^2+36a_0^2)\log ~y +(10y^2+63a_0^2)\log(y^2+9a_0^2)\Big]\Bigg\},
\nd
where $M$ is not a constant and is a function of the radial coordinate $r$ given via $f(r)$ defined in eqn (2.17) of \cite{mia2}. Using $f(r)$, one may show that $M(r)$ asymptotes to 
zero in Region 3. The integration is performed from $r = a_0$ instead of $r = 0$ to avoid singularities\footnote{In fact in the presence of a black hole, 
the $r = 0$ region will be covered 
by the horizon $r_h$ so this will not be the issue when thermal limit is considered as we will see later. 
In the absence of a black hole, but in the presence of the deformation parameter, this will again not be 
an issue.}.  
We have also defined $\kappa = {r^2 + 9a_0^2\over r^2 + 6a_0^2}$ such that $\kappa = 1 + {3a_0^2\over r^2}$ for $r > a_0$; and 
$\kappa = {3\over 2} - {r^2\over 12 a_0^2}$ for $r < a_0$. When $r = a_0$, $\kappa = {10\over 7}$, a constant factor. In effect $\kappa$ ranges from 1.5 to 1 for $r$ ranging from 
$r = 0$ to $r = \infty$, although the decrease is not monotonous. 

The integration of the NS B-field over the two-cycle $S^2$ in \eqref{dileshwar} is now important. What two-cycle should we choose? One choice would be the resolution two-cycle 
($\theta_2, \phi_2$). However we could equally choose ($\theta_1, \phi_1$), because for $r \ge a_0$, there is not much difference between the two two-cycles.    
We are therefore
interested in the integral along $(\theta_1, \phi_1)$. This is straightforward, with only a small subtlety arising with the first term in \eqref{b2} resulting in an 
improper $\theta_1$ integral which must be regulated by taking a cutoff near the poles of the 2-sphere and sending it to zero after integrating. The final result is:
\bg\label{sep2014}
\int_{S^2} B_2 ~=  -4\pi b_1+4\pi b_3 + \frac{3 g_s^2 M N_f}{
 4 \pi} \Bigg\{1 + \log \left(r^2 + 9 a^2 \right) \left[-1 + \log \left( \sin \frac{\theta_2}{2} \right) \right] \Bigg\}. \nonumber\\
\nd

\begin{figure}[h]\label{intb2}
  \centering
    \includegraphics[width=0.7\textwidth]{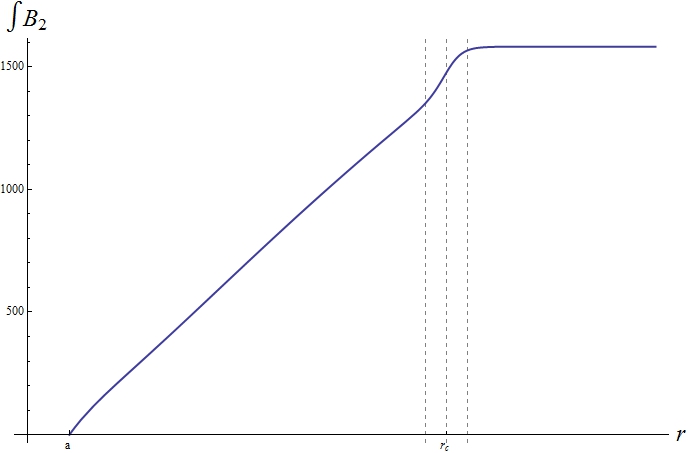}
  \caption{$\int B_2$ as a function of $r$. It grows steadily in Region 1, then grows faster in Region 2 before reaching a constant value in Region 3, where all fluxes get shut off. The dashed lines indicate Region 2, centered around the cutoff radius $r_c$. The size of Region 2 is controlled by the rate at which the effective $M(r)$ is switched off and depends on the exact distribution of the anti-D5 charges.}
\label{intb2}
\end{figure}

\noindent The first and last term will vanish in Region 3, due to the effective number of 5-branes, $M(r)$, being gradually turned off in Region 2. The middle term only has $M(r)$ dependence in its derivative and will therefore plateau in Region 3 as depicted in {\bf fig \ref{intb2}}. 
This asymptotic value determines the location of our theory along the Klebanov-Witten fixed surface in the UV. We also want our gravity description to be dual to a confining theory in 
the IR, which determines the value of $\int B_2$ at minimal radius. Choosing a different initial value corresponds to choosing the number of colours in the UV to \emph{not} be a integer 
multiple of $M$. This results in the IR theory to be different from the confining $SU(M)$ that we are interested in.

Let us now discuss the behavior of the dilaton $\Phi$ in our model.
In the absence of the seven-branes the dilaton would be a constant dictated by our choice of string coupling, but the presence of D7-branes introduces a logarithmic correction which 
can be computed from the monodromy around the D7-branes. As discussed earlier, this is the expected behavior in Regions 1. The result is:
\bg\label{dila1}
e^{-\Phi} = {1\over g_s} -\frac{N_f}{8\pi} ~{\rm log} \left(r^6 + 9a_0^2 r^4\right) - 
\frac{N_f}{2\pi} {\rm log} \left({\rm sin}~{\theta_1\over 2} ~ {\rm sin}~{\theta_2\over 2}\right).
\nd
We can fix the last term to be a constant by either choosing a particular slice with fixed values of $\theta_1, \theta_2$ away from the location of the D7-branes 
or by taking an average over the base of the conifold. In either case, since we are ultimately interested in the radial dependence of the dilaton, we will simply absorb this constant 
shift of the dilaton into our choice of $g_s$. Note that, since we are using Ouyang embedding in Region 1 \cite{mia2, ouyang}, the D7-branes do not go all the way down to $r = 0$. However 
notice that generically the radial logarithmic correction ensures that $e^{-\Phi}$ reaches zero at finite radius, leading to a Landau pole for both couplings. 
This behavior makes Region 3 a necessary component of the model.

\begin{figure}[h]\centering \includegraphics[width=0.7\textwidth]{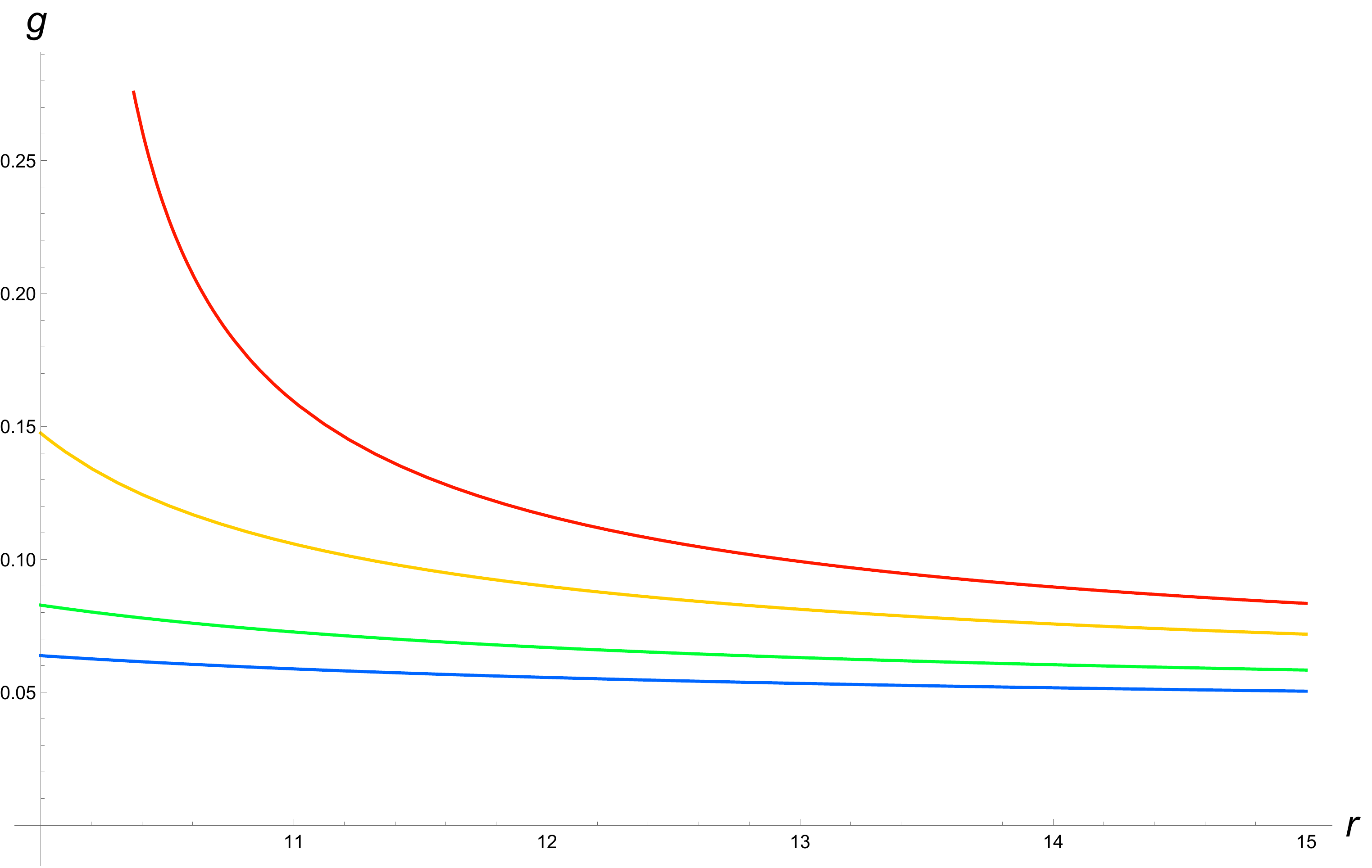} \caption{Behavior of the coupling in Region 3 for different choices of asymptotic UV value with the cutoff radius $r_c=10$} \label{region3} \end{figure}

In Region 3, the behavior of the dilaton changes from increasing logarithmically to a decay asymptoting to a finite value in the UV. The functional form of the beta function can be 
determined from F-theory \cite{mia2}:
\bg\label{dila2}
\beta_{e^{-\Phi}} = \frac{{\bf C}_0}{r^2(r^{3/2}-1)},
\nd
with the constant ${\bf C}_0$
depending on the details of the UV cap that is attached. Integrating the beta function gives the behavior shown in {\bf fig \ref{region3}}. 
We see that the dilaton asymptotes to a constant value in the UV. Combined with the constant $\int B_2$ in Region 3, this stops the RG flow at a UV fixed point located somewhere on 
the Klebanov-Witten fixed surface for the corresponding $SU(N + M)\times SU(N + M)$ theory.

\subsection{The RG Flow at Strong Coupling \label{rgfull}}

\begin{figure}[h]
  \centering
    \includegraphics[scale=0.3]{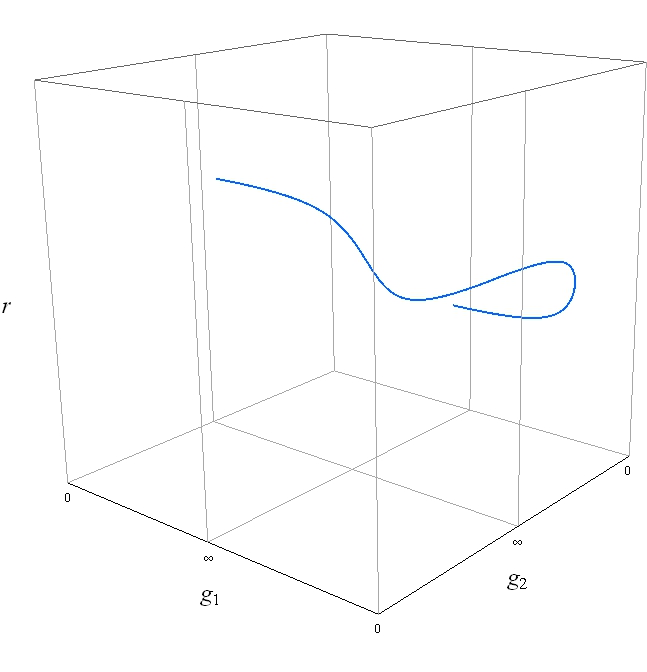}
  \caption{The RG flow through a single Seiberg duality. When $g_2$ diverges, we switch to the Seiberg dual description where $g_2$ now represents the dual coupling. This new coupling 
decreases, while $g_1$ diverges instead, so we dualize it instead. This process continues till we hit the end of the cascade.}
\label{rg2cycles}
\end{figure}

We are now in a position to describe the entire RG flow of the theory from UV to IR at strong coupling. Using \eqref{dileshwar}, the two couplings may be represented in terms of 
supergravity variables as:
\bg\label{alujen}
&& {8\pi^2\over g_1^2} = {e^{-\Phi}\over 2}\left[{1\over 2} + {1\over 2\pi}\left(\int_{S^2} {B}_2 \mod 2\pi\right)\right] \nonumber\\
&& {8\pi^2\over g_2^2} = {e^{-\Phi}\over 2}\left[{3\over 2} - {1\over 2\pi}\left(\int_{S^2} {B}_2 \mod 2\pi\right)\right]. \nd 
Aside from the logarithmic correction from introducing fundamental flavors, the flow in Region 1 is essentially the same as the KS scenario. We always interpret $g_1$ as the coupling of 
the \emph{lower} rank gauge group. Since Seiberg duality changes which of the gauge groups has higher rank, the interpretation of which $g_i$ belongs to which group keeps changing every time $\int B_2$ changes by $4 \pi^2$. Each cycle of $\int B_2$ starts with a divergent $g_1$ and finite $g_2$. As $\int B_2$ grows $g_2$ increases, while $g_1$ decreases. Eventually $g_2$ diverges, indicating the need to Seiberg dualize that part of the gauge group. Upon doing this the higher-rank gauge group becomes the lower-rank one so its gauge coupling is now represented by $g_1$ instead, which is again divergent, while $g_2$ has the same value that $g_1$ had at the end of the previous cycle. We can thus connect two consecutive $4 \pi^2$ cycles of $\int B_2$ smoothly as shown in {\bf fig \ref{rg2cycles}}. Continuing this process we recover a smooth looking flow. 

Note that the divergence of the gauge couplings does not indicate any special features in the geometry, since the quantities $g_1$ and $g_2$ do not have a clear physical interpretation in the gravity description. Indeed as emphasized in \cite{0505153}, the gravity description is oblivious to the duality cascade. Any measure of the number of degrees of freedom on the gravity 
side will indicate a smooth decrease rather than a sequence of sudden jumps from Seiberg duality, as is the case in the field theory at low coupling.

\begin{figure}[h]
  \centering
    \includegraphics[scale=0.4]{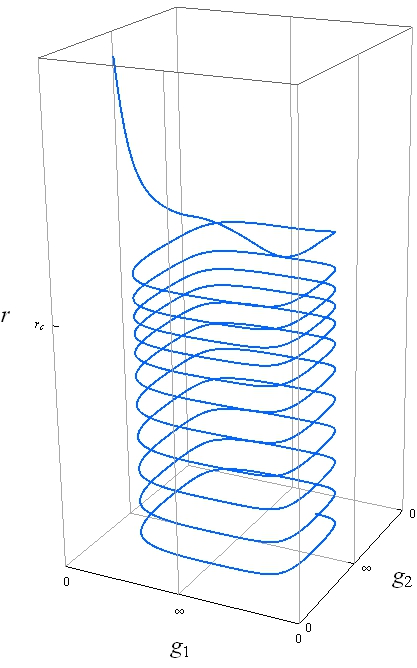}
  \caption{RG flow near ther cutoff radius $r_c$. The last few Seiberg dualities happen at closer energies, due to the rapid change of $M(r)$ in Region 2 and therefore $\int B_2$. 
At higher energies, in Region 3, the couplings asymptote to their UV values.}
\label{region2}
\end{figure}

Region 2 is similar qualitatively to Region 1, except for the behavior of the B-field due to the change in the effective number of five-branes $M(r)$. 
This increases the rate at which we need to perform Seiberg dualities as shown in {\bf fig \ref{region2}}, while also decreasing the changes in the gauge group rank upon each duality. 
Eventually the B-field asymptotes to a fixed value, the gauge group ranks become equal and the two gauge couplings stop flowing relative to each other\footnote{There is still a walking
RG flow due to the fundamental flavors. For details see \cite{chempot}.}. 
Thus region two serves as a smooth interpolation between the cascading behavior of the KS model and the asymptotically conformal behavior of Region 3.

In Region 3, the only flow is due to the behavior of the dilaton. As the dilaton asymptotes to a fixed but finite value, so do the gauge couplings and the theory becomes conformal, 
although not asymptotically free. However we can also choose a functional form for the dilaton such that the gauge couplings asymptotes to zero. This way, in the language of 't Hooft 
coupling, the theory becomes conformal, but in the language of gauge coupling, the theory becomes asymptotically free.  

\begin{figure}[h]
  \centering
    \includegraphics[scale=0.4]{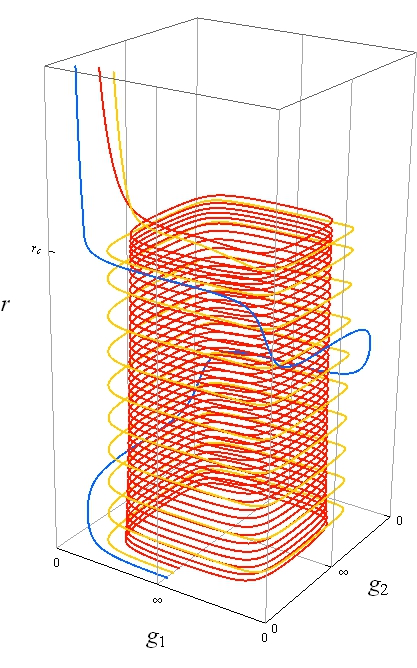}
  \caption{Complete RG flow from UV to IR for three different asymptotical UV values and the same $M(r)$. The weaker coupled flow (blue) flow slower than the stronger coupled flows (red). 
All three flows end in confinement as $g_1$  diverges for the last time.}
\label{rgsameM}
\end{figure}

Note that for each choice of gauge couplings keeping the number of colors in the UV we have a different dual geometry, with a different choice of asymptotic value of the dilaton and cutoff 
radius at which we attach Region 3. To compare flows for several initial choices of coupling we need to either have a different cutoff radius, or rescale $M(r)$ so that each theory 
undergoes the same number of Seiberg dualities between the Higgsing energy scale and the IR. In the former case, as shown in  
{\bf fig \ref{rgsameM}}, we see that weaker coupling results in slower RG flow. 

\begin{figure}[h]
  \centering
    \includegraphics[scale=0.4]{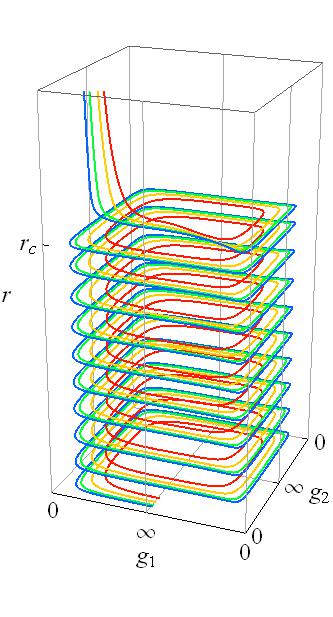}
  \caption{Complete RG flow from UV to IR for four different asymptotical UV values with appropriately scaled $M(r)$ for each flow.}
\label{rgplotfull}
\end{figure}

This is expected, since we know from the gauge theory description that at weak coupling, the flow will slow down near the Seiberg fixed points. The gravity analysis does not extend to that regime, where $\mathcal{O}(\frac{1}{g_s M}, \frac{1}{g_s N})$ corrections are expected to alter the shape of the flow, but the overall slowing of the flow is evident. If we instead rescale $M(r)$ the flows for different choices of UV couplings look more similar, but each flow corresponds to a different numbers of colors in the dual gauge theory 
as shown in {\bf fig \ref{rgplotfull}}. 

\section{Towards Bulk Viscosity from the Gravity Dual \label{bulky}}

In the previous two sections we studied the stability and the RG flows in our model. Our discussion was mostly in the zero temperature limit, as no black hole was inserted in the 
gravity dual. In the presence of a black hole, 
thermal effects in gauge theory do not change any of the earlier conclusions. For example thermal stability can be inferred from an analysis simlar to what was done in section 
\ref{stability} (see also \cite{aalok}). Similarly, thermal beta functions resemble the ones discussed in section \ref{Rogi}. The latter aspect has also been studied in 
section 2.3 of \cite{chempot}.

Of course new phenomena do arise from thermal effects. Many of them have been studied earlier in \cite{fep, mia2, melting, chempot, mesonic, aalok, patra, ramalata}. In the following 
section, we will study another interesting thermal effect, the bulk viscosity. One distinguising feature of bulk viscosity, compared to say the shear viscosity, 
is the {\it necessity} of non-conformality since in the conformal limit the bulk viscosity vanishes. Our study will involve both conformal and non-conformal regimes, and we will be 
able to confirm the vanishing of bulk viscosity in the conformal case. For the non-conformal case we will be able to lay out the calculational scheme using the UV complete model and determine the precise form of the bulk viscosity, including the 
ratio of the bulk viscosity to entropy density in terms of a function that depends on the details of the UV completion of the 
model. We relegate a more detailed study for \cite{richard}.  

\subsection{Setup of System and Metric}

We begin a complete top-down analysis of bulk viscosity of type IIB supergravity with a black hole.  We begin with veilbeins:
\bg\label{arrange} 
&& e_6 = \frac{re^{-A}}{3}\left(d\psi+\cos~\theta_1~d\phi_1+\cos~\theta_2~d\phi_2\right) \nonumber\\
&& e_7 = \frac{re^{-A}}{\sqrt{6}}\left[\cos\left(\frac{\psi}{2}\right)d\theta_1+\sin\left(\frac{\psi}{2}\right)\sin~\theta_1~d\phi_1\right] \nonumber\\
&& e_8 = \frac{re^{-A}}{\sqrt{6}}\left[\sin\left(\frac{\psi}{2}\right)d\theta_1-\cos\left(\frac{\psi}{2}\right)\sin~\theta_1~d\phi_1\right] \nonumber\\
&& e_9 = \frac{re^{-A}}{\sqrt{6}}\left[\cos\left(\frac{\psi}{2}\right)d\theta_2+\sin\left(\frac{\psi}{2}\right)\sin~\theta_2~d\phi_2\right] \nonumber\\
&& e_{10} = \frac{re^{-A}}{\sqrt{6}}\left[\sin\left(\frac{\psi}{2}\right)d\theta_2-\cos\left(\frac{\psi}{2}\right)\sin~\theta_2~d\phi_2\right] \nonumber\\
&& e_1 = e^{A+B}dt, ~~e_2 = e^{A}dx, ~~ e_3 = e^{A}dy, ~~ e_4 = e^{A}dz, ~~ e_5 = e^{-A-B}dr.  
\nd
From these veilbeins, we can build all the elements of our type IIB supergravity.  Let's analyze the components of the veilbeins.  The first four coordinates, $e_1$ to $e_4$,  
describe Minkowski space, albeit with a warp factor $e^{A}$.  Also, a feature of this model is the presence of a black hole which manifests itself as a factor of $e^{B}$ on the $dt$ 
veilbein.  The other six coordinates, $e_5$ to $e_{10}$,  
depict a conifold, warped as well by the warp factor $e^{-A}$.  Another component of the black hole is attached to the veilbein for the radial coordinate.  Using the 
vielbeins \eqref{arrange} we begin to build the type IIB background that we need, beginning with the metric:
\bg\label{metric}
ds_{10}^2 &=& \eta^{ab}e_a e_b = - e^{2A+2B}dt^2+e^{2A}dx^2+e^{2A}dy^2+e^{2A}dz^2+e^{-2A-2B}dr^2 \\
&& + \frac{r^2 e^{-2A}}{6}\left[\sum_{i = 1}^2 (d\theta_i^2 +\sin^2~\theta_i~d\phi_i^2)\right] + \frac{r^2e^{-2A}}{9}(d\psi +\cos~\theta_1~d\phi_1 +\cos~\theta_2~d\phi_2)^2.\nonumber 
\nd
Note that the internal space in \eqref{metric} is not a warped deformed cone as one might have expected. This choice is not just for analytical simplicity, but is governed by 
two underlying facts. One, the deformation parameter that would capture the far IR regimes of the dual gauge theory is now covered by the horizon and therefore for $r > r_h$, we basically 
see a conifold geometry. Two, the analysis presented in this and the next two subsections will concentrate mostly on the conformal regimes of our geometry and therefore a conifold rather 
than a defomed conifold will be more relevant. Thus, for $r > r_h$, all the three regions, namely Regions 1, 2 and 3, with internal conifold metric would be a sufficiently good approximation.  
 
From the metric \eqref{metric}
we can build the various gravitational components such as the Ricci scalar $R$ and the Ricci tensor $R_{\mu\nu}$.  Next, we build the five-form flux due to the D3-branes. 
We equate the 4-potential to the volume of the Minkowski coordinates as:
\begin{equation}
C_4 = e^{-B}{\bf vol}_4 \equiv e^{-B}\bigwedge_{n=1}^{4}e_n = e^{4A}dt\wedge dx\wedge dy\wedge dz.
\end{equation}
Note that we have inserted a corrective factor of $e^{-B}$.  We will be assuming that only the metric becomes non-extremal due to the black hole, and we will investigate the effects this has 
on the various type IIB flux components.  
From our definition of $C_4$, we can simply build the five-form flux, making it self dual, which is a consequence of type IIB supergravity:
\begin{equation}
F_5 = (1+\ast_{10})dC_4,
\end{equation}
where, $\star_{10}$ is the Hodge Dual with respect to the ten dimensional metric. Next, we discuss the complex three-form flux $G_3$ on the conifold. For this, first we  
combine the six veilbeins \eqref{arrange} into three complex one-forms in the following way:
\bg\label{compv}
E_1 = e^{B}e_5 + ie_6, ~~~~~ E_2 = e_7 + ie_8, ~~~~~ E_3 = e_9 + ie_{10}, 
\nd
where we have again inserted a corrective factor of $e^B$ in $E_1$ to remove the non-extremal effect of the black hole.  Using these one-forms, we can construct our three-form flux as:
\bg\label{G3}
G_3=\frac{9M}{4r^3}~E_1\wedge\left(E_2\wedge\bar{E}_2-E_3\wedge\bar{E}_3\right), 
\nd
which by construction is a non-ISD three-form, and becomes ISD once the black-hole is removed. 
The parameter $M$ here is the number of $D5$-branes, that is, the number of bifundamental flavors that we encountered earlier. We are in Region 1, so $M(r) \approx M$, and we won't worry 
about the anti-D5 branes right now (they will appear soon). 
The factor of $\frac{9}{4}$ is determined by insisting that the extremal correction to the warp factor (found in section \ref{ActEqns}) matches that found by Klebanov and Strassler 
in \cite{KS}.  These $D5$-branes wrap around one of the two-cycles $(\theta_1,\phi_1)$ and fill the four Minkowski coordinates. 
Because the two-cycles are compact, these $D5$-branes act as fractional $D3$-branes. For now, our axio-dilaton will be constant:
\bg\label{dilato}
\tau = i/g_s.
\nd
This can be changed later by turning on $N_f$ $D7$-branes as one may infer from \eqref{dila1} (see also \cite{ouyang, fep, mia2}).

\subsection{Action and Equations of Motion in the Conformal Limit \label{ActEqns}}

Our aim in this section is to determine the precise functional forms for $e^A$ and $e^B$ using the background ansatze for the metric \eqref{metric}, $G_3$ flux \eqref{G3} and the 
axio-dilaton \eqref{dilato}. To proceed, we start with the type IIB supergravity action as given in \cite{GKP}:
\bg\label{iibaction}
S_{IIB} = \frac{1}{2\kappa_{10}^2}\int{d^{10}x\sqrt{-g}\left[R-\frac{\partial\tau\cdot\partial\bar{\tau}}{2({\rm Im}~\tau)^2}-\frac{G_3\cdot\bar{G}_3}{12{\rm Im}~\tau}
-\frac{F_5^2}{4\cdot5!}\right]} + \frac{1}{8i\kappa_{10}^2}\int{\frac{C_4 \wedge G_3 \wedge \bar{G}_3}{{\rm Im}~\tau}}. \nonumber\\ 
\nd
As discussed above, we are in Region 1 and therefore mostly analyzing the IR regime of our theory. The ansatze for $G_3$ flux is \eqref{G3}, and in the limit when 
$M = 0$, we have switched off non-conformality altogether. This is then equivalent to Region 3 instead where the $G_3$ flux vanishes, alongwith vanishing $N_f$, the seven-brane 
degrees of freedom. This means the conformal theory is in the regime where the sizes of Regions 1 and 2 are vanishing, and the physics is captured completely by Region 3 
only\footnote{Where the gauge group is $SU(N) \times SU(N)$. We could also take $SU(N+M) \times SU(N+M)$, as one would expect from Region 3, but this is just a redefinition of the number 
of colors. As such, in the following sections, we would like to keep $M$ solely as a signal of broken conformal invariance.}. 
For this case, we expect: 
\bg\label{beom}
dF_5 = \frac{G_3 \wedge \bar{G}_3}{\bar{\tau}-\tau} = 0.
\nd
In order for the left hand side of (\ref{beom}) to be non-zero, we must impose by hand the self duality of the five-form flux:
\begin{equation}
F_5 = (1+\ast_{10})dC_4.
\end{equation}
The remaining set of equations are the Einstein equations which in general may be expressed with source terms coming from $G_3$ and $F_5$ fluxes in the following way:
\bg
&& R_{\mu\nu} = -g_{\mu\nu}\left[\frac{G_3\cdot\bar{G}_3}{48{\rm Im}~\tau}+\frac{F_5^2}{8\cdot5!}\right]+\frac{F_{\mu abcd}F_{\nu}^{abcd}}{4\cdot4!} \label{MinkEins}\\
&& R_{mn} = -g_{mn}\left[\frac{G_3\cdot\bar{G}_3}{48{\rm Im}~\tau}+\frac{F_5^2}{8\cdot5!}\right]+\frac{F_{mabcd}F_{n}^{abcd}}{4\cdot4!}
+\frac{G_m^{bc}\bar{G}_{nbc}}{4{\rm Im}~\tau}+\frac{\partial_m\tau\partial_n\bar{\tau}}{2|{\rm Im}~\tau|^2}. \label{RadEins}
\nd
In our system, we have two undetermined scalar functions, $A(r)$ and $B(r)$ appearing in the metric ansatze \eqref{metric}.  
Using (\ref{MinkEins}), we can solve for $B(r)$.  First notice that:
\bg\label{notice}
R_t^t-R_x^x=0, 
\nd
and therefore inserting our metric ansatze \eqref{metric} in \eqref{notice}, we get a simple differential equation for $B(r)$:
\begin{equation}
\frac{d^2B}{dr^2}+\frac{dB}{dr}\left(\frac{5}{r}+2\frac{dB}{dr}\right)=0. 
\end{equation}
The solution to which is:
\begin{equation}
B(r) = \frac{1}{2}{\rm log}\left(c_2 + \frac{c_1}{r^4}\right).
\end{equation}
To solve for the constants $c_1$ and $c_2$, we use the two boundary conditions, one, that $g\equiv e^{2B}$ vanishes at the black hole horizon $r_h$ and two,  
that $e^{2B}\rightarrow 1$ at the conformal boundary $r\rightarrow\infty$. These conditions are satisfied by:
\bg\label{bitta}
B(r) = \frac{1}{2}{\rm log}\left(1-\frac{r_h^4}{r^4}\right) \equiv \frac{1}{2}{\rm log}\left[g(r)\right], 
\nd
Where we refer to the function $g(r)$ as the black hole factor. With this definition in hand, we can move on to the five-form equation of motion 
(\ref{beom}) which allows us to find $A(r)$.  The explicit equation when $M = N_f = 0$ is:
\begin{equation}
\frac{d^2A}{dr^2}+\frac{dA}{dr}\left(\frac{5}{r}-4\frac{dA}{dr}\right)=0, 
\end{equation}
whose solution may be written as:
\begin{equation}
A(r) = -\frac{1}{4}{\rm log}\left(\frac{L^4}{r^4}\right) \equiv -\frac{1}{4}{\rm log}\left[h(r)\right].
\end{equation}
Here, $L^4=\frac{2\pi g_sN}{4}$, where $N$ is the number of $D3$-branes. The function $h(r)$ is what we referred to as the warp factor of the system, as it controls, among other things, 
the factor by which the first four coordinates are warped from Minkowski space at a given value of the AdS radius $r$. With this, our system is completely defined. We may now go on to 
using the system and the AdS/CFT duality to calculate interesting and relevant quantities on the field theory side. 
The inclusion of the black hole in the system, as expected, gives the field theory a temperature depending on the black hole radius $r_h$.

\subsection{Diagonal Perturbations and Bulk Viscosity in the Conformal Limit \label{coiofto}}

We wish to calculate the bulk viscosity using the Kubo formula:
\begin{equation}
\zeta = \frac{1}{18}\lim_{\omega\rightarrow 0}\lim_{\vec{k}\rightarrow 0}\frac{1}{\omega}\int d^4x e^{i\omega t-i\vec{k}\cdot\vec{x}}\left\langle\left[T_{ii}(t,\vec{x}),T_{jj}(0,\vec{0})\right]\right\rangle
\end{equation}
Here, the sum over $i,j \in \left\lbrace x,y,z\right\rbrace$ is implied. Because nothing in the system depends upon any one given spatial direction, we have that the only $\vec{k}$ dependence if the above expression is in the complex exponential. The independence of the system on Minkowski spatial directions means the system has an $SO(3)$ symmetry, 
implying $T_{xx}=T_{yy}=T_{zz}$. So our simplified Kubo formula is:
\begin{equation}
\zeta = \lim_{\omega\rightarrow 0}\frac{1}{2\omega}\int d^4x e^{i\omega t}\left\langle\left[T_{xx}(t,\vec{x}),T_{xx}(0,\vec{0})\right]\right\rangle.
\end{equation}
We see then that the bulk viscosity is related to the retarded propagator:
\bg\label{retpro}
&& \zeta ~ = ~ -\lim_{\omega\rightarrow 0}\frac{{\rm Im}(G_R(\omega,0)}{\omega} \nonumber\\
&& G_R(\omega,\vec{k}) ~= ~ -i\int d^4x e^{i\omega t-i\vec{k}\cdot\vec{x}} \theta(t) \left\langle\left[T_{xx}(t,\vec{x}),T_{xx}(0,\vec{0})\right]\right\rangle.
\nd
One immediate advantage of expressing the bulk viscosity in terms of the retarded propagator is its connection to the Schwinger-Keldysh propagator. 
Following \cite{herson}, we can relate the retarded propagator to the Schwinger-Keldysh propagator as:
\begin{equation}
G_{11}^{SK}(\omega,\vec{k})={\rm Re}\left[G_R(\omega,\vec{k})\right] + i\coth\left(\frac{\omega}{2T}\right){\rm Im}\left[G_R(\omega,\vec{k})\right].
\end{equation}
This will then allow us to express the bulk viscosity as:
\bg\label{zeta}
\zeta=-\frac{1}{2T}\lim_{\omega\rightarrow 0}{\rm Im}\left[G_{11}^{SK}(\omega,0)\right]. 
\nd
Here we have used the fact that $\coth\left(\frac{\omega}{2T}\right)\rightarrow\frac{\omega}{2T}$ for small $\omega$.  The Schwinger-Keldysh (SK) propagator can be derived by considering the field theory action on a Schwinger-Keldysh contour.  From this analysis, we obtain the following definitions:
\bg\label{oppro}
iG_{11}^{SK}(t,\vec{x})~=~ - \frac{\delta^2\ln[Z_{CFT}(\phi_1,\phi_2)]}{\delta\phi_1(t,\vec{x})\delta\phi_1(0,0)} ~= ~\left\langle\mathcal{T}\mathcal{O}_1(t,\vec{x})\mathcal{O}_1(0,0)\right\rangle,
\nd 
where $\mathcal{T}$ is the time ordering symbol.  The operator product in \eqref{oppro} may now be given the following meaning.  
If we choose $\phi_1(t,\vec{x})$ to be the boundary value of $X(t,r)$, i.e the graviton perturbation along the $x$ direction, 
then this will mean that $\mathcal{O}_1(t,\vec{x})=T_{xx}(t,\vec{x})$.  
The AdS-CFT conjecture states that we can replace $Z_{CFT}$ with $Z_{sugra}$ defined using the type IIB action as $e^{iS_{IIB}}$.  Therefore, we need only expand the type $IIB$ supergravity action to second order in the graviton perturbation, Fourier transform the action, and then apply the above functional derivative to obtain an expression for $G_{11}^{SK}(\omega,0)$, which we can then plug into the definition for the bulk viscosity. Similar procedure is discussed for the shear viscosity in \cite{fep}. 
Following \cite{herson}, when we eventually find our perturbations, they will take the following form:
\bg\label{phipm}
\phi_{\pm}(r,\omega)=\phi\left(r,\pm\frac{i|\omega|}{4\pi T}\right),  
\nd
where $\omega$ is as defined earlier. Using $\phi_{\pm}$, we can now define $\phi_1$ more accurately as:
\bg\label{phi1}
\phi_1(r,\omega)=a_0\left[\phi_{+}(r,\omega)-e^{\frac{\omega}{T}}\phi_{-}(r,\omega)\right], 
\nd
where $a_0$ is a constant (and not to be confused with the bare resolution parameter defined earlier).
With these tools in hand, we need only find the functional form of the relevant perturbation.  We perturb the veilbeins ($e_1, ..., e_5$) in \eqref{arrange} in the following way:
\bg\label{perturb}
&& e_k ~=~ e^{A}\left[1+X(r,t)\right]dx_k \nonumber\\
&& e_1 ~=~ e^{A+B}\left[1+T(r,t)\right]dt \nonumber\\
&& e_5 ~= ~ e^{-A-B}\left[1+R(r,t)\right]dr, \nd  
with ($e_6, ..., e_{10}$) remaining unchanged as \eqref{arrange}. We have also taken $k = 2, 3, 4$ and defined ($dx_2, dx_3, dx_4$) as ($dx, dy, dz$) respectively in \eqref{perturb}. The 
above deformation captures the essence of bulk viscosity: if we change the overall size of the system, any {\it resistance} we encounter will signal the presence 
of a {\it bulk} viscosity.  

We need all three of the perturbations $T(r,t)$, $X(r,t)$ and $R(r,t)$ in \eqref{perturb}
because the equations of motion we will derive are heavily coupled with respect to these perturbations.  We plug these veilbeins into our system components and then 
into our equations of motion and expand these equations to the first order in the perturbations. For example the vielbeins \eqref{perturb} induce 
a metric fluctuation $\delta g_{\alpha\beta}$, 
such that the EOM for the fluctuation to first order becomes:
\bg\label{einsto}
\frac{\partial R_{\mu\nu}}{\partial g_{\alpha\beta}} -{1\over 2}\left(g_{\mu\nu}{\partial R\over \partial g_{\alpha\beta}} + g^\alpha_\mu g^\beta_\nu R\right) 
= {\partial T_{\mu\nu} \over \partial g_{\alpha\beta}}, \nd
where $T_{\mu\nu}$ is the energy momentum tensor that come from the background fluxes, and ($R_{\mu\nu}, R$) are the usual Ricci tensor and Ricci scalar. The higher order contributions from 
vielbein fluctuations can be easily computed, but we will not do so here. However,  
before we lay out the equations to solve them, we Fourier transform our metric perturbations in the following way:
\bg\label{forier} 
T(t,r)&=&\int_{-\infty}^{+\infty} e^{it\omega}\widetilde{T}\left(r,\omega\right)d\omega \nonumber\\
X(t,r)&=&\int_{-\infty}^{+\infty}  e^{it\omega}\widetilde{X}\left(r,\omega\right)d\omega \nonumber\\
R(t,r)&=&\int_{-\infty}^{+\infty}  e^{it\omega}\widetilde{R}\left(r,\omega\right)d\omega, 
\nd
where note that although $\Gamma_i \equiv$ ($T, X, R$) are real perturbations, the Fourier components 
$\widetilde{\Gamma}_i \equiv$ ($\widetilde{T}, \widetilde{X}, \widetilde{R}$) can have complex pieces. Thus generically we can express the 
Fourier coefficients as:
\bg\label{forcoe}
\widetilde{\Gamma}_i ~ = ~ {\rm Re} ~\widetilde{\Gamma}_i + i {\rm Im}~\widetilde{\Gamma}_i, \nd
and the existence of the non-zero imaginary piece, at least for the $\widetilde{X}$ Fourier component, will signal the presence of a bulk viscosity. On the other hand, the 
reality of $\Gamma_i$ will at least imply:
\bg\label{reality} 
\widetilde{\Gamma}_i(r, \omega) = \widetilde{\Gamma}^\ast_i(r, -\omega), \nd
where $\ast$ is the complex conjugation. One may also impose a more global integral condition, but if \eqref{reality} is satisfied then it is more apparent. Note that \eqref{reality}
also implies that we will need odd powers of $\omega$ to counteract the $\ast$ action.  

We now analyze all the supergravity equations of motion using the Fourier decomposition given in \eqref{forier}. Since we don't have three-form fluxes (we are as in Region 3), the supergravity
EOMs consist of the Einstein and the five-form flux equations. The Einstein equation \eqref{einsto} for the $tt$ component can be written as:
\bg\label{tt} 
\widetilde{T}''+ \widetilde{T}'\left(\frac{5}{r}+2B'\right)-\left(\widetilde{T}'+3\widetilde{X}'\right)A'-\widetilde{R}'\left(A'+B'\right)
+\omega^{2}e^{-4A-4B}\left(3\widetilde{X}+ \widetilde{R}\right)=0, \nonumber\\ \nd
where the derivative is wrt to the radial direction $r$. Using \eqref{forcoe} the above equation can be split into two equations for the real and the imaginary parts of $\widetilde{\Gamma}_i$. 

The second is the graviton fluctuation along ($x, y, z$) directions. However since we expect the energy momentum tensors along the three spatial directions to be identical, we can study only the
$xx$ Einstein equation. In terms of Fourier components, this is given by:
\bg\label{xx} 
\widetilde{X}''+ \widetilde{X}'\left(\frac{5}{r}+2B'\right)-\left(\widetilde{T}'+3\widetilde{X}'\right)\left(A'+B'\right)- \widetilde{R}'A'+\omega^{2}e^{-4A-4B}\widetilde{X} = 0, \nd  
where as before we can decompose this in terms of real and complex pieces. The other components, namely $yy$ and $zz$ graviton fluctuations, will take exactly the same form as \eqref{xx}.
On the other hand, the $rr$ graviton fluctuation will be different and is given by:
\bg\label{rr} 
\left(\widetilde{T} + 3\widetilde{X}\right)''- \widetilde{R}'\left(\frac{5}{r}+2B'\right) + \widetilde{T}'\left(A'+2B'\right) + 3\widetilde{X}'A' + \widetilde{R}'\left(A'+B'\right) + 
\omega^2 e^{-4A-4B}\widetilde{R} = 0. \nonumber\\ \nd 
Again the above equation is linear in the Fourier components, and therefore the complex components of the equation will take similar form. This looks like generic, and so the complex parts would
solve identical equations. Can this be different? A hint may come from the $rt$ fluctuation of the graviton which exists
because of the time dependence of the perturbations. The equation takes the following form:
\bg\label{rt}
\frac{d}{dt}\left[3 {X}'- 3 {X} B'- {R}\left(\frac{5}{r}-2A'\right)\right] = 0, \nd 
where note that we wrote this in terms of ($X, R$) and not in terms of the Fourier components ($\widetilde{X}, \widetilde{R}$). One implication of this is that we can rewrite \eqref{rt} without the time 
derivative as:
\bg\label{rt2} 
3 {X}'- 3 {X} B'- {R}\left(\frac{5}{r}-2A'\right) = c_0, \nd
$c_0$ is a time-independent function (here it could simply be a function of $r$). 
However in terms of the Fourier components, the only solution for  $c_0$ is that it vanishes identically.  
This means that the 
real and the complex parts of \eqref{rt2} would again be identical. 
Finally, the Bianchi identity for $F_5$ 
leads to the following equation:
\bg\label{df5} 
\left(\widetilde{T} + 3\widetilde{X}\right)''+\left(\widetilde{T} + 3\widetilde{X}\right)'\left(\frac{5}{r}-4A'\right) - 4A'\widetilde{R}'=0. \nd 
This array of equations seems daunting, given that there are an excess of equations with respect to variables (5 to 3), but there is a consistent solution.  
With careful combinations of \eqref{tt} + 3\eqref{xx}, \eqref{rr} and \eqref{df5}
and taking inspiration from the shear viscosity calculation in \cite{fep}, we postulate that: 
\bg\label{Xr} 
\widetilde{X}(r, \omega) = e^{2aB(r)}, \nd 
where $a$ is now a function of $\omega$ (which could be complex) and $B(r)$ is given in \eqref{bitta}. Plugging \eqref{Xr} in the set of equations \eqref{tt}, \eqref{xx}, \eqref{rr} and \eqref{df5}, 
we arrive at the following consistent solution for the other two Fourier components:
\bg\label{xrtsol} 
\widetilde{T}(r, \omega) = \left(1-{2\over a}\right)e^{2aB(r)}, ~~~
\widetilde{R}(r, \omega) = 2(2a-1)\left(e^{-2B(r)}-1\right)e^{2aB(r)}. \nd
The quantity $a$ appearing above, as mentioned earlier, is a function of $\omega$ and can be expressed in terms of $L$ and the horizon radius $r_h$ as:
\bg\label{chabat} 
a(\omega) = 1 + \frac{L^4\omega^2}{8r_h^2} = 1 + \frac{\omega^2}{8\pi^2 T^2} = 1 - 2\left(\pm\frac{i\vert\omega\vert}{4\pi T}\right)^2.
\nd
Note that these are solutions that require that we solve the equations in the limit in which $r=r_h$, exactly as in the calculation for shear viscosity.  The last expression for $a$ is in terms of 
$\gamma \equiv \frac{i\vert\omega\vert}{4\pi T}$, which will allow us to express future solutions in terms of the same power 
of the black hole that comprises the solution for the shear diagonal perturbation (see eq (3.175) in \cite{fep}):
\begin{equation}
\phi(r,\omega) = e^{\gamma B(r)}. 
\end{equation}
However, for our case 
note that although the solution for $\widetilde{X}(r,\omega)$ 
depends explicitly on $\gamma$, the solution is also real, meaning that it will eventually lead to a bulk viscosity solution of $\zeta=0$.  Generically however, in the 
set of equations \eqref{tt}, \eqref{xx}, \eqref{rr} and \eqref{df5}, the real and the imaginary parts of the fluctuations ($X, T, R$) satisfy identical equations.
For such a case we can either have Im $\widetilde{\Gamma}_i = 0$ in \eqref{forcoe} for $a$ satisfying \eqref{chabat}, or: 
\bg\label{ghushi}
\widetilde{\Gamma}_i(r, \omega) ~ = ~ \left(1 \pm i\sum_{n=0}^\infty b_{in} \omega^{2n+1}\right){\rm Re}~\widetilde{\Gamma}_i\left(r, \vert\omega\vert\right), \nd
to ensure the reality of the fluctuations \eqref{forier} using \eqref{reality}, as ${\rm Re}~\widetilde{\Gamma}_i$ is expressed in even powers of $\omega$. However the 
evaluation of bulk viscosity requires us to go to the limit $\omega \to 0$ according to \eqref{zeta}. In this limit the imaginary part of \eqref{ghushi} vanishes.
This is as it should be: in the conformal limit, we expect the bulk viscosity to vanish.  The purpose of finding the exact form of this solution is twofold: one, 
the conformal solution confirms a bulk viscosity of zero and two, the form of the conformal solution will act as a base upon which we build any non-conformal corrections.  Any non-conformal corrections that lead to a non-zero bulk viscosity \textit{must} lead to a perturbation $\widetilde{X}(r,\omega)$ that has a non-zero imaginary part (the choice of the imaginary piece is subtle, as we will clarify 
later).

\subsection{Towards Bulk Viscosity in the Non-Conformal Limit}

We now add the effects of the $D5$-branes to the system by setting $M \neq 0$.  This will lead to corrections to the metric, the warp factor and the black hole factor which are all controlled
by the small parameter:
\begin{equation}
\epsilon ~= ~ \frac{81g_sM^2}{8L^4}~ = ~ \frac{3g_s M^2}{2\pi N}.
\end{equation}
We begin with the corrections to the metric, as it will affect all the other components of the system.  Besides the corrections to the warp factor, the metric picks up explicit corrections of 
order $\epsilon$ via a resolution parameter\footnote{This is not quite the resolution parameter that we encountered earlier in section \ref{dilatone}. In 
section \ref{dilatone} we took a warped {\it resolved} conifold so as to study the UV behavior. This is the {\it brane} side i.e the gauge theory side of the problem.
Here, as we concentrate only on the IR behavior (i.e integrate out the anti-D5 brane DOFs) and as we are in the gravity dual, 
we take a conifold so that the bare resolution 
parameter vanishes. As such we can write $a^2(r) = {\cal O}(\epsilon)$. In the 
following we will be able to determine the ${\cal O}(\epsilon)$ corrections.}  
$a^2(r)$:
\bg\label{eye}
&& e_5 = e^{-A-B}\sqrt{\frac{r^2+6a(r)^2}{r^2+9a(r)^2}} dr \nonumber\\
&& e_6 = \frac{re^{-A}}{3}\sqrt{\frac{r^2+9a(r)^2}{r^2+6a(r)^2}}\left(d\psi+\cos~\theta_1 d\phi_1+\cos~\theta_2 d\phi_2\right) \nonumber\\
&& e_7 = \frac{re^{-A}}{\sqrt{6}}\left[\cos\left(\frac{\psi}{2}\right)d\theta_1+\sin\left(\frac{\psi}{2}\right)\sin~\theta_1 d\phi_1\right] \nonumber\\
&& e_8 = \frac{re^{-A}}{\sqrt{6}}\left[\sin\left(\frac{\psi}{2}\right)d\theta_1-\cos\left(\frac{\psi}{2}\right)\sin~\theta_1 d\phi_1\right] \nonumber\\
&& e_9 = e^{-A}\sqrt{\frac{r^2+6a(r)^2}{6}}\left[\cos\left(\frac{\psi}{2}\right)d\theta_2+\sin\left(\frac{\psi}{2}\right)\sin~\theta_2 d\phi_2\right] \nonumber\\
&& e_{10} = e^{-A}\sqrt{\frac{r^2+6a(r)^2}{6}}\left[\sin\left(\frac{\psi}{2}\right)d\theta_2-\cos\left(\frac{\psi}{2}\right)\sin~\theta_2 d\phi_2\right]. 
\nd 
We see that putting the $M$ $D5$-branes on one of the two 2-spheres has caused an asymmetry quantified by the resolution parameter.  Our assumption is that $a(r)^2=\mathcal{O}(\epsilon)$ and has no terms that are zeroth order in $\epsilon$.  This can be confirmed by plugging the metric into the equations of motion.  Furthermore, we note that we must have $a(r)=0$ in the limit that $r_h=0$, in order to recover the IR conifold solution (with $M(r) \approx M$).  
We can assume that inserting $D5$-branes into a {\it non-extremal} system will affect the warp factor in some way.  We quantify this effect using the function $G(r)$:
\begin{equation}\label{watbits}
h(r) = h_0(r)+\epsilon G(r),
\end{equation}
so that the corrections are of order $\epsilon$ and higher. We will also use the shorthand to express the resolution parameter in the following way:
\begin{equation}\label{trose}
a(r)^2 = \epsilon F(r)
\end{equation}
in order to put all non-extremal effects on the same footing.
Additionally, we can also create the combination of the contracted Einstein equations to allow us to find an exact solution for the black hole factor:
\begin{equation}\label{bhole}
g(r) = 1+4r_h^4\int^r\frac{dy}{y^3\left[y^2+9\epsilon F(y)^2\right]}.
\end{equation}
We turn now to Einstein's equations.  The $tt$ and $xx$ equations are structurally the same, with an extra factor of $g_0 = 1-\frac{r_h^4}{r^4}$ in front of the $tt$ equation.  So, the only real different equations are the $xx$ equation:
\bg\label{eqnxx}
&& \left(\frac{d^2G}{dr^2}-\frac{3}{r}\frac{dG}{dr}-\frac{36}{r^3}\frac{dF}{dr}+\frac{4}{r^2}\right)\left(1-\frac{r_h^4}{r^4}\right)+\frac{72 F}{r^4}\left(1+\frac{r_h^4}{r^4}\right)\nonumber\\
&& ~~~~~~~~~~ +\frac{2 r_h^4}{r^4}\left(\frac{dG}{dr}+\frac{1}{r^2}+288r^2\int^r{\frac{F(y) dy}{y^7}}\right)=0,
\nd
with $G(r)$ and $F(r)$ are as given above in \eqref{watbits} and \eqref{trose} respectively, and the $rr$ equation:
\bg\label{eqnrr}
&& \left(\frac{d^2G}{dr^2}-\frac{3}{r}\frac{dG}{dr}+\frac{30}{r^2}\frac{d^2F}{dr^2}-\frac{66}{r^3}\frac{dF}{dr}+\frac{4}{r^2}\right)\left(1-\frac{r_h^4}{r^4}\right)
+\frac{72 F}{r^4}\left(1+\frac{r_h^4}{r^4}\right)\nonumber\\
&& ~~~~~~~~~~~ -\frac{2 r_h^4}{r^4}\left(2\frac{dG}{dr}+\frac{1}{r^2}+288r^2\int^r{\frac{F(y) dy}{y^7}}+\frac{144}{r^4}F\right)=0.
\nd
The above set of equations seem formidable, but we can form the much simpler combinations $rr + xx$, i.e \eqref{eqnxx} + \eqref{eqnrr} to get the following equation:
\begin{equation}
\frac{d^2G}{dr^2}-\frac{3}{r}\frac{dG}{dr}+\frac{15}{r^2}\frac{d^2F}{dr^2}-\frac{51}{r^3}\frac{dF}{dr}+\frac{72}{r^4}F+\frac{4}{r^2}=0 \label{rrmxx}
\end{equation}
and the opposite combination, $rr - xx$ i.e \eqref{eqnrr} $-$ \eqref{eqnxx} to get the following equation:
\begin{equation}
-15\left(1-\frac{r_h^4}{r^4}\right)\left(\frac{d^2F}{dr^2}-\frac{1}{r}\frac{dF}{dr}\right)+\frac{144 r_h^4 F}{r^6} + 576 r_h^4\int^r{\frac{F(y) dy}{y^7}}
+ \frac{2r_h^4}{r^4}-4r\frac{dG}{dr}=0. \label{rrpxx}
\end{equation}
From each of these, we can solve for $G(r)$.  Solving \eqref{rrmxx}, we can express $G(r)$ using $F(r)$ in the following way:
\begin{equation}
G(r)=\int^r y^3 dy \left[C_1 -\int^y dx\left({15\over x^5} {d^2F\over dx^2} - {51\over x^6} {dF\over dx} + {72 F(x)\over x^7} + {4\over x^5}\right)\right] + C_2, \label{G1}
\end{equation}
where $C_1$ and $C_2$ are constants.
On the other hand, solving \eqref{rrpxx} yields another functional form for $G(r)$ in terms of $F(r)$ in the following way:
\begin{equation}
G(r) = \frac{1}{4}\int^r dy \left[-\frac{15 y^3}{r_h^4}\left(1-\frac{r_h^4}{y^4}\right)\left(\frac{d^2F}{dy^2}-\frac{dF}{dy}\right)+\frac{144 F(y)}{y^3} + \frac{2}{y}\right]dy 
+ \widetilde{C}_1, \label{G2}
\end{equation}
where $\widetilde{C}_1$ is another constant. 
We proceed to find $F(r)$ by equating the right hand sides of \eqref{G1} and \eqref{G2}, simplifying and taking two derivatives, we get a second order differential equation for $f(r)=\frac{dF}{dr}$:
\begin{equation}
\frac{d^2f}{dr^2}-\frac{1}{r}\frac{df}{dr}+\frac{f}{r^2}=\frac{2}{15g_0}\frac{dg_0}{dr},
\end{equation}
where note that the constants $C_1, C_2$ and $\widetilde{C}_1$ get automatically removed so that we have a second-order differential equation without any extra constants; and 
$g_0=1-\frac{r_h^4}{r^4}$ is the conformal black hole factor.  This can be solved to find:
\begin{equation}
f(r) = K_1r\ln{r}+K_2r+\frac{2r}{15}\operatorname{dilog}(g_0),
\end{equation}
were $K_1$ and $K_2$ are constants. We integrate once more, to get:
\begin{equation}
F(r) = r^2\left\{\widetilde{K}_1+\widetilde{K}_2\ln{r}+\frac{1}{30}\left[\frac{1}{2}\operatorname{dilog}(g_0)-\ln{(g_0)}+\frac{r_h^2}{r^2}\ln{\left(\frac{r^2-r_h^2}{r^2+r_h^2}\right)}\right]\right\},
\end{equation}
where we have repackaged the $K_i$ constants into two other constants $\widetilde{K_1}$ and $\widetilde{K_2}$.   
We require that $F(r)$ obey certain limits.  We need $F(r)$ to disappear in the limit $r_h\rightarrow 0$ and we need $F(r)$ to be finite in the limit $r\rightarrow r_h$.  
The first limit is so that our result matches the extremal result, i.e  there is no ${\cal O}(\epsilon)$ correction to the resolution parameter.  
The second limit ensures that calculations we do later to find the bulk viscosity do not diverge.  The first limit then results in the conditions:
\begin{equation}
\lim_{r_h\rightarrow 0}\widetilde{K}_1=0,\quad \lim_{r_h\rightarrow 0}\widetilde{K}_2=0.
\end{equation}
Since $\widetilde{K}_1$ and $\widetilde{K}_2$ are dimensionless, we must have that $\widetilde{K}_i=a_i\left(\frac{r_h}{L}\right)^{b_i}$, where the $b_i>0$.  
The simplest case is $a_i=0$ for $i=1,2$.  Taking this into account, we can now plug the full result for $F(r)$ back into \eqref{G1}, and perform the integrals to get a final solution for $G(r)$:
\begin{equation}
G(r) = \ln{r}+\frac{1}{5}\left[\ln(g_0) - \frac{r_h^2}{r^2}\ln\left(\frac{r^2-r_h^2}{r^2+r_h^2}\right)-\frac{1}{8}\operatorname{dilog}(g_0)\right],
\end{equation}
which behaves well in the limit $r \to r_h$ as one would expect. 
With these, our unperturbed non-conformal system is fully defined to ${\cal O}(\epsilon)$ that we seek here.  We have:
\begin{eqnarray}\label{fatur}
&& a(r)^2=\frac{\epsilon r^2}{30}\left[-\ln(g_0)+\frac{r_h^2}{r^2}\ln\left(\frac{r^2-r_h^2}{r^2+r_h^2}\right)+\frac{1}{2}\operatorname{dilog}(g_0)\right] \\
&& h(r) = \frac{L^4}{r^4}\left\{1 + \epsilon \left[\ln{r}+\frac{1}{5}\left(\ln(g_0) - \frac{r_h^2}{r^2}\ln\left(\frac{r^2-r_h^2}{r^2+r_h^2}\right)
-\frac{1}{8}\operatorname{dilog}(g_0)\right)\right]\right\}. \nonumber
\end{eqnarray}
We move now to setting up the system of equations that will allow us to solve for the $\mathcal{O}(\epsilon)$ corrections to the metric perturbations. This will come from both 
the Einstein's EOMs as well as the flux equations. In the language of the Fourier modes, we expect a set of equations that would take the following order by order expansion in 
the small parameter $\epsilon$:
\bg\label{seteqn}
\mathbb{F}_0(r, \omega) + \epsilon~ \mathbb{F}_1(r, \omega) + \epsilon^2~ \mathbb{F}_2(r, \omega) + {\cal O}(\epsilon^3) ~ = ~ 0, \nd
where $\omega$ is the Fourier frequencies, and $\mathbb{F}_0(r, \omega)$ for example will denote the $\epsilon$ independent i.e the conformal results \eqref{tt}, \eqref{xx}, 
\eqref{rr}, \eqref{rt}, and \eqref{df5}. We also expect all the parameters involved here are now expressed as expansions in $\epsilon$, i.e:
\bg\label{epexp}
h &=& h_0 + \epsilon h_1 + \epsilon^2 h_2 + {\cal O}(\epsilon^3) \nonumber\\ 
g &=& g_0 + \epsilon g_1 + \epsilon^2 g_2 + {\cal O}(\epsilon^3) \nonumber\\
A &=& A_0 + \epsilon A_1 + \epsilon^2 A_2 + {\cal O}(\epsilon^3) \nonumber\\                                                     
B &=& B_0 + \epsilon B_1 + \epsilon^2 B_2 + {\cal O}(\epsilon^3), \nd
with the subscript 0 denoting the conformal results, $h \equiv e^{-4A}$ is the warp-factor involved in describing the background 
and $g \equiv e^{2B}$ is the black-hole factor. Similarly the 
resolution factor, as we studied above, has the expansion $a^2 = f_0 + \epsilon f_1(r) + {\cal O}(\epsilon^2)$, with $f_0 = 0$ for the conifold case that we consider here and 
$f_1(r)$ may be derived from \eqref{fatur}. 
For the
Fourier components of the metric fluctuations \eqref{forier},                   
we simply make the 
substitutions:
\bg\label{jutamaro} 
&&{\widetilde T}(r,\omega) = T_0(r,\omega)+\epsilon \left[T_{11}(r,\omega) + iT_{12}(r,\omega)\right] + {\cal O}(\epsilon^2) \nonumber\\
&& {\widetilde R}(r,\omega) = R_0(r,\omega)+\epsilon \left[R_{11}(r,\omega) + i R_{12}(r,\omega)\right] + {\cal O}(\epsilon^2)\nonumber\\ 
&& {\widetilde X}(r,\omega) = X_0(r,\omega)+\epsilon \left[X_{11}(r,\omega) + iX_{12}(r,\omega)\right] + {\cal O}(\epsilon^2),  
\end{eqnarray} 
where $T_0$, $X_0$ and $R_0$ are the Fourier modes satisfying the set of equations \eqref{tt}, \eqref{rr}, \eqref{xx}, \eqref{rt2} and \eqref{df5} 
whose solutions are \eqref{Xr} and 
\eqref{xrtsol}. 
As mentioned earlier, they are 
all real. 
The goal now is to find the real and the imaginary components ($T_{11}, X_{11}, R_{11}$) and ($T_{12}, X_{12}, R_{12}$) respectively. 
We will exploit the fact that, to any order in $\epsilon$, the 
set of equations \eqref{epexp} should yield:
\bg\label{yield} 
\mathbb{F}_0(r, \omega) ~ \equiv ~ 0, ~~~~~ \mathbb{F}_1(r, \omega) ~ \equiv ~ 0, ~~~~~ \mathbb{F}_2(r, \omega) ~ \equiv ~ 0, \nd
so that the number of variables in the expansion \eqref{jutamaro} should at least match up with the number of equations. Needless to say, 
the set of equations $\mathbb{F}_0(r, \omega) ~ \equiv ~ 0$
are the conformal equations \eqref{tt}, \eqref{rr}, \eqref{xx}, \eqref{rt2} and \eqref{df5}.
 
Again, our equations are the $tt$, $xx$, $rr$ and $rt$ components of Einstein's equations as well as the Bianchi identities for the $F_5$ and now the $G_3$ flux and we will 
concentrate only to first order in $\epsilon$ here. 
For the real components, 
the left hand side of the new equations $\mathbb{F}_1 \equiv 0$,  
will be identical to the left hand side of the set of the equations \eqref{tt}, \eqref{rr}, \eqref{xx}, \eqref{rt2} and \eqref{df5}. 
The right hand side of these equations will no longer be zero, but will be source terms that depend on $T_0(r)$, $X_0(r)$, $R_0(r)$ and now $F(r)$ and $G(r)$. The source terms will be the terms 
from the left hand side of the set of equations \eqref{tt}, \eqref{rr}, \eqref{xx}, \eqref{rt2} and \eqref{df5}, 
where the factor of $\epsilon$ comes from something other than the perturbations, namely the warp factor $h(r)=e^{-4A(r)}$ and the black hole factor $g(r)=e^{2B(r)}$, 
as well as from naturally occuring $\mathcal{O}(\epsilon)$ terms from the $G_3$ contributions to the Einstein equations and Bianchi identity. There could also be contributions 
from the smeared anti five-brane of Regions 2 and 3 that we have ignored so far (see equation (2.27) of \cite{mia2} for complete 
details\footnote{There is a small typo in eq (2.27) of \cite{mia2}: The numerator in the first term should be $\partial_{(m}\bar\tau \partial_{n)}\tau$ instead of 
$\partial_{(m}\partial_{n)}\tau$.}). We can quantify this by adding additional sources as:   
\bg\label{danach} 
\Delta^{(\alpha)}(r, \omega) \equiv 0 + \epsilon\left[\Delta^{(\alpha)}_{11}(r, \omega) + i\Delta^{(\alpha)}_{12}(r, \omega)\right] + {\cal O}(\epsilon^2),\nd
where to zeroth order in $\epsilon$ all sources have already been taken into account earlier, and $\alpha = 1, 2, 3$ correspond to $T, R$ and $X$ fluctuations respectively.

To the first order in $\epsilon$, 
the $tt$ Einstein equation then gives us the following equation connecting the real parts ($T_{11}, X_{11}, R_{11}$) to the sources and the real components ($T_0, X_0, R_0$) of 
the set of equations \eqref{tt}, \eqref{rr}, \eqref{xx}, \eqref{rt2} and \eqref{df5}:
\bg\label{tta}
&& T_{11}''+T_{11}'\left(\frac{5}{r}+2B_0'\right)-\left(T_{11}'+3X_{11}'\right)A_0'-R_{11}'\left(A_0'+B_0'\right)+\frac{\omega^{2}h_0}{g_0^2}\left(3X_{11}+ R_{11}\right) \nonumber\\
&&= - 2B_1'T_0' + \left(T_0'+3X_0'\right)A_1'
- \frac{\omega^{2}(g_0h_1-2g_1h_0)}{g_0^3}\left(3X_0+R_0\right) \nonumber\\ 
&& ~~~~~~~~~~~~~~~~~~~ + R_0'\left(A_1'+B_1'\right) + \left(\frac{1+g_0}{g_0 r^2}\right)T_0 + \Delta^{(1)}_{11},
\nd
where ($X_0, R_0, T_0$) are given by \eqref{Xr} and \eqref{xrtsol}. Note that, in the absence of the source term $\Delta^{(1)}_{11}$, the equation \eqref{tta} 
is linear in terms of the fluctuations, and the inhomegeneity in the equation should only appear from the additional source $\Delta^{(1)}_{11}$. Unless mentioned otherwise, this will be the 
case for all the equations below.    
The other terms appearing in \eqref{tta} can be derived from the supergravity background and are given by:
\bg\label{chukka}
&& g_0 = 1-\frac{r_h^4}{r^4}, ~~~~B_1 = -\frac{18r_h^4}{g_0^2}\int^r{\frac{dx ~F(x)}{x^7}}\\
&& A_0 = -\frac{1}{4}\ln\left({L^4\over r^4}\right), ~~~~ B_0 = \frac{1}{2}\ln{\left(1-\frac{r_h^4}{r^4}\right)}, ~~~~ 
g_1 = -36r_h^4\int^r{\frac{dx~F(x)}{x^7}}\nonumber\\
&& h_1 = -4A_1 h_0, ~~~~ h_0 = \frac{L^4}{r^4}, ~~~~ A_1 = -\frac{1}{4}\left[{G(r) \over h_0}\right]. \nonumber
\nd
On the other hand, the imaginary part of the $tt$ Einstein equation for the Fourier components may now be expressed in the following way:
\bg\label{ttb}
T_{12}''+T_{12}'\left(\frac{5}{r}+2B_0'\right)-\left(T_{12}'+3X_{12}'\right)A_0'-R_{12}'\left(A_0'+B_0'\right)+\frac{\omega^{2}h_0}{g_0^2}\left(3X_{12}+ R_{12}\right) 
= \Delta^{(1)}_{12}, \nonumber\\  \nd
where the coefficients are defined above. Note that the 
terms the LHS of \eqref{ttb} is similar to the equation \eqref{tt} for the real or the imaginary pieces of the Fourier components for the conformal case. 

Let us now go to the real part of the $xx$ Einstein equation. The form is somewhat similar to the $tt$ Einstein equation \eqref{tta}, but certain details differ. The equation is:
\bg\label{xxa}
&& X_{11}''+X_{11}'\left(\frac{5}{r}+2B_0'\right)-\left(T_{11}'+3X_{11}'\right)\left(A_0'+B_0'\right)-R_{11}'A_0'+ \frac{\omega^{2}h_0}{g_0^2}X_{11} \nonumber\\
&& = -2B_1' X_0' + \left(T_0'+3X_0'\right)\left(A_1'+B_1'\right) - \frac{\omega^2(g_0h_1-2g_1h_0)}{g_0^3}X_0 \nonumber\\
&&~~~~~~~~~~ + R_0'A_1' + \left(\frac{1+g_0}{g_0r^2}\right)X_0 + \Delta^{(2)}_{11},
\nd
where all the coefficients are defined earlier in \eqref{chukka}. As before, this is only to the first order in $\epsilon$, and thus mixes with ($X_0, T_0, R_0$). 
The imaginary part 
of the equation to this order takes the following form: 
\bg\label{xxb}
X_{12}''+X_{12}'\left(\frac{5}{r}+2B_0'\right)-\left(T_{12}'+3X_{12}'\right)\left(A_0'+B_0'\right)-R_{12}'A_0' + \frac{\omega^{2}h_0}{g_0^2}X_{12} = \Delta^{(2)}_{12}, \nonumber\\ \nd
which as before is similar to \eqref{xx} for the real or the imaginary pieces of the Fourier components. 

The other spatial components of Einstein equations, namely $yy$ and $zz$ components, are identical to \eqref{tta} because of isometry so we will concentrate on the $rr$ 
equation. The real part of the equation takes the following form:
\bg\label{rra}
&& \left(T_{11}+3X_{11}\right)''-R_{11}'\left(\frac{5}{r}+2B_0'\right)+T_{11}'\left(A_0'+2B_0'\right)+3X_{11}'A_0'+R_{11}'\left(A_0'+B_0'\right)\nonumber\\ 
&& + \frac{\omega^2h_0}{g_0^2}R_{11} = 2B_1'R_0' - T_0'\left(A_1'+2B_1'\right)-3X_0'A_1'-R_0'\left(A_1'+B_1'\right) \nonumber\\
&&~~~~~~~~~ - \frac{\omega^2(g_0h_1-2g_1h_0)}{g_0^3}R_0+\left(\frac{3g_0 - 1}{g_0r^2}\right)R_0  + \Delta^{(3)}_{11}, \nd
and as expectedly, the imaginary part takes similar form as the real or the imaginary parts of \eqref{rr}, namely:
\bg\label{rrb}
&& \left(T_{12}+3X_{12}\right)''-R_{12}'\left(\frac{5}{r}+2B_0'\right)+T_{12}'\left(A_0'+2B_0'\right)+3X_{12}'A_0' \nonumber\\
&& ~~~~~~~~ +R_{12}'\left(A_0'+B_0'\right) 
+ \frac{\omega^2h_0}{g_0^2}R_{12} = \Delta^{(3)}_{12}. \nd 
The next set of equations appear from the $rt$ components of the Einstein equation. Again this equation would exist because of the time dependence of the fluctuations. The real part 
now takes the following form:
\bg\label{rta}
3X_{11}'-3X_{11} B_0'- R_{11}\left(\frac{5}{r}-2A_0'\right) - 3 X_0 B_1' - 2 A_1'R_0 = {\rm Re} ~{\bf C}, \nd
where ${\bf C}$ is in general a function of $\Delta^{(\alpha)}_{ab}$. Such a term would be absent for the conformal case \eqref{rt} as one would expect. 
In fact existence of this influences the 
imaginary part of the $rt$ fluctuation equation in the following way:
\bg\label{rtb}
3X_{12}'-3X_{12} B_0'- R_{12}\left(\frac{5}{r}-2A_0'\right) = {\rm Im}~{\bf C}. \nd
Looking at \eqref{rtb} we are tempted again to compare with \eqref{rt}. 
There are however two possibilities now: 

\vskip.1in 

\noindent $\bullet$ One, when ${\rm Im}~{\bf C} = \Delta^{(\alpha)}_{12} = 0$, then the imaginary parts of the fluctuation equations, \eqref{ttb}, \eqref{xxb}, \eqref{rrb} and \eqref{rtb}
match with the imaginary parts of the 
fluctuation equations \eqref{tt}, \eqref{xx}, \eqref{rr} and \eqref{rt}. 

\vskip.1in 

\noindent $\bullet$ Two, when ${\rm Im}~{\bf C}$ and $\Delta^{(\alpha)}_{12}$ are non-vanishing, then the imaginary parts of fluctuation equations 
\eqref{ttb}, \eqref{xxb}, \eqref{rrb} and \eqref{rtb}
in general do not 
match with either the real or the 
imaginary parts of the fluctuation equations \eqref{tt}, \eqref{xx}, \eqref{rr} and \eqref{rt}. 

\vskip.1in 

\noindent Note that the behavior of ${\rm Re}~{\bf C}$ and $\Delta^{(\alpha)}_{11}$ do not effect our discussion because the 
real parts of the equations \eqref{tta}, \eqref{xxa}, \eqref{rra} and 
\eqref{rta} are very different from the real parts of \eqref{tt}, \eqref{rr}, \eqref{xx} and 
\eqref{rt}. We will discuss more on this later.

Finally, let us go to the flux equations. First, is the EOM coming from the three-form flux $G_3$. However, 
we do not need to concern ourselves with the $G_3$ equation at this point because $\mathcal{O}(\epsilon)$ corrections to the metric perturbations result in $\mathcal{O}(\epsilon^2)$ 
corrections to the equation. Thus the $G_3$ EOM will start changing the results only to ${\cal O}(\epsilon^2)$. Similarly,  
the axion EOM will not contribute anything because we are not taking
the $g_sN_f$ backreactions into account. We expect the sources \eqref{danach} to only affect the Einstein's 
equations\footnote{Note that, as long as there are no induced fluxes on the anti five-brane sources in Regions 2 and 3 $-$ quantified here by \eqref{danach} $-$ we expect the $G_3$ and 
axion EOMs to have no contributions to this order.}. 

The second is then the five-form EOM. This will contribute as before, with the real part of the equation taking the following form:   
\bg\label{df5a}
&& \left(T_{11} + 3X_{11}\right)''+\left(T_{11} + 3X_{11}\right)'\left(\frac{5}{r}-4A_0'\right)-4A_0'R_{11}'\nonumber\\  
&& ~~~~~~~~ = 4A_1' \left(T_0+3X_0\right)' + 4A_1'R_0'-\frac{4R_0}{r^2} + {\rm Re}~{\bf D}, \nd
where ${\bf D}$ is another function of the sources $\Delta^{(\alpha)}_{ab}$ similar to ${\bf C}$ above. This imples, as before, the LHS of the  
imaginary part of the equation takes the form similar to the real or the imaginary parts of the equation \eqref{df5} encountered 
earlier:
\bg\label{df5b}
\left(T_{12} + 3X_{12}\right)''+\left(T_{12} + 3X_{12}\right)'\left(\frac{5}{r}-4A_0'\right)-4A_0'R_{12}' = {\rm Im}~{\bf D}. \nd   
We now have all the equations we need to solve to first order in $\epsilon$ for the fluctuations given in \eqref{jutamaro}. For bulk viscosity, it is important that the imaginary 
parts of the fluctuations in \eqref{jutamaro} are non-zero. To analyze this let us consider the sources \eqref{danach} to be non-zero. The precise functional
form for $\Delta^{(\alpha)}_{ab}$ is 
now necessary to
relate the set of equations \eqref{ttb}, \eqref{xxb}, \eqref{rrb}, \eqref{rtb} and \eqref{df5b} 
to the imaginary parts of the set of equations \eqref{tt}, \eqref{xx}, \eqref{rr}, \eqref{rt} and \eqref{df5} respectively. In the absence of the precise knowledge of 
$\Delta^{(\alpha)}_{ab}$, and the fact that the LHS of all the imginary parts of the fluctuations match with the ones for the conformal case, lead us to propose the 
following possible 
solutions to the fluctuations \eqref{jutamaro}:
\bg\label{nore2n25}
&& \widetilde{T}_\pm(r, \omega) ~ = ~ \left(1 \pm i\epsilon \sum_{n = 0}^\infty p_n \omega^{2n-1}\right) T_0\left(r, \vert\omega\vert\right) 
+  \epsilon T_{11}\left(r, \vert\omega\vert\right) + {\cal O}(\epsilon^2) \nonumber\\
&& \widetilde{R}_\pm(r, \omega) ~ = ~ \left(1 \pm i\epsilon \sum_{n = 0}^\infty q_n \omega^{2n-1}\right) R_0\left(r, \vert\omega\vert\right) 
+  \epsilon R_{11}\left(r, \vert\omega\vert\right) + {\cal O}(\epsilon^2) \nonumber\\ 
&& \widetilde{X}_\pm(r, \omega) ~ = ~ \left(1 \pm i\epsilon \sum_{n = 0}^\infty f_n \omega^{2n-1}\right) X_0\left(r, \vert\omega\vert\right) 
+  \epsilon X_{11}\left(r, \vert\omega\vert\right) + {\cal O}(\epsilon^2), \nd
where ($X_0, R_0, T_0$) are the values \eqref{Xr} and \eqref{xrtsol} for the conformal case studied earlier, ($p_n, q_n, f_n$) are real functions of $r$,   
and ($T_{11}, R_{11}, X_{11}$) are the solutions to the 
real parts of the fluctuation equations, i.e the set of equations \eqref{tta}, \eqref{xxa}, \eqref{rra}, \eqref{rta}, and \eqref{df5a}. Clearly this is an over-determined system, but 
as for the conformal case, we expect solutions to exist\footnote{The appearance of $\vert\omega\vert$ on the RHS of \eqref{nore2n25} implies that they are even powers of $\omega$, as 
should be clear from the EOMs governing the fluctuations.}.  

The way we have constructed the solutions in \eqref{nore2n25}, they satisfy the reality condition \eqref{reality} and are functions of $r$ and $\omega$. We will eventually have to consider 
the limit when 
$\omega$ approaches zero. In this limit, the imaginary parts of \eqref{nore2n25} take the following form:
\bg\label{thelate}
&&{\rm Im}~{\widetilde T}_\pm(r, \omega) = \pm\left[{p_0 \epsilon \over \omega} + p_1 \epsilon \omega + {\cal O}(\omega^2)\right]T_0\left(r, \vert\omega\vert\right) 
 \to  \pm{p_0 \epsilon\over \omega} T_0\left(r, \vert\omega\vert\right) \nonumber\\
&&{\rm Im}~{\widetilde R}_\pm(r, \omega) = \pm\left[{q_0 \epsilon \over \omega} + q_1 \epsilon \omega + {\cal O}(\omega^2)\right]R_0\left(r, \vert\omega\vert\right) 
 \to  \pm {q_0 \epsilon\over \omega} R_0\left(r, \vert\omega\vert\right)\\
&&{\rm Im}~{\widetilde X}_\pm(r, \omega) = \pm \left[{f_0 \epsilon \over \omega} + f_1 \epsilon \omega + {\cal O}(\omega^2)\right]X_0\left(r, \vert\omega\vert\right) 
 \to  \pm {f_0 \epsilon\over \omega} X_0\left(r, \vert\omega\vert\right), \nonumber \nd
where we see that there is an interesting simplification: the result {\it only} depend on the functional forms of 
$p_0(r), q_0(r)$ and $f_0(r)$. All other functions $p_i(r), q_i(r)$ and $f_i(r)$ for $i \ne 0$ are irrelevant for the specific computation that we aim for here! Additionally, as we 
shall soon see, it is in fact only the functional form for $f_0(r)$ that will eventually be required in the bulk viscosity computation\footnote{There is a subtlety here. When $f_0$ is a constant, 
the bulk viscosity vanishes despite the existence of an imaginary piece to the $\widetilde{X}_{\pm}$ fluctuation. Thus having an imaginary piece to the $\widetilde{X}_{\pm}$
fluctuation is a necessary but not a sufficient condition for the existence of a non-zero bulk viscosity.}. 
This amazing simplification is of course 
only for our specific computation, and for all other transport coefficients, we will require the full knowledge of the functions $p_n(r), q_n(r)$ and $f_n(r)$ unless of course we 
go to the $\omega \to 0$ limit. Note however that, although all the values in \eqref{thelate} seem to blow-up in this limit, the bulk viscosity will be {\it finite} 
in this limit. Needless to say, 
such a solution can {\it only} exist in the non-conformal limit where we have a way to introduce a tunable parameter $\epsilon$. In the 
conformal limit, and as we saw from \eqref{ghushi}, a non-zero imaginary piece to the fluctuation cannot exist. 

Before moving forward, let us clarify one issue related to $\gamma$ defined earlier. 
For the conformal case we used a parameter $\gamma \equiv {i\vert\omega\vert\over 2\pi T}$ in \eqref{Xr} and \eqref{xrtsol} to determine the fluctuations. The same $\gamma$ appears in eqn (3.174) of \cite{fep} for the determination of shear viscosity. For terms with {\it even} powers of $\vert\omega\vert$, the reality condition \eqref{ghushi} 
is naturally satisfied. However for the bulk viscosity computation, if we express our result using the parameter $\gamma$, how is the reality condition \eqref{ghushi} satisfied now?

The answer turns out to be the way we have expressed \eqref{nore2n25} and \eqref{thelate}: we are in principle not required to use parameter $\gamma$. However if we instead use the technique of \cite{fep} $-$ discussed for shear viscosity $-$ then the reality issue will come back. Note that
in \cite{fep}, the fluctuation $\phi(r, t)$ was defined as: \bg\label{krii} \phi(r, t) = \int_0^\infty d\omega g^\gamma F(r, \gamma) \varphi(\omega), \nd
where $g(r)$ is the black-hole factor. This makes sense as positive energy i.e $\omega > 0$ case was studied in \cite{fep}. If we want to consider all energies, we have to just add a
complex conjugate piece to \eqref{krii}. This way $\phi(r, t)$ will be real\footnote{Adding the complex conjugate implies $\vert\omega\vert \to \omega$ in all the expressions in the 
shear viscosity computation of \cite{fep}. Thus all analysis, using only positive energy fluctuation, remain unchanged, as expected. This can also be seen from eqn (3.2) in \cite{katpet} where the physical fluctuation was taken to be complex, which could be made real by adding a complex conjugate. This implies no change in the analysis as emphasized above.}.  

Coming back, the way one would now go about computing the bulk viscosity from the complex fluctuations \eqref{nore2n25} and \eqref{thelate} is to express the total type IIB action 
completely in the language of $\widetilde{T}_\pm(r, \omega), \widetilde{R}_\pm(r, \omega)$ and $\widetilde{X}_\pm(r, \omega)$, 
much like eqn (3.170) of \cite{fep}\footnote{This further implies that the $r$ integral would run from $r_h$ to $r_c$, the cut-off radius. This cut-off radius $r_c$ is similar to 
the cut-off radius $r_c$ that we encountered in section \ref{Rogi}.}, but now expressed in terms of all the 
three Fourier components ($\widetilde{T}_\pm, \widetilde{R}_\pm, \widetilde{X}_\pm$).  
Note that the action remains real but the imaginary pieces, essential for the bulk viscosity computation, appear solely 
from the Fourier components (as was also the case for the shear viscosity computation in \cite{fep}). 
For the specific case here, we start by defining
$\widetilde{X}_1(r, -\omega)$ as certain combination of modes $\widetilde{X}_+(r, -\omega)$ and $\widetilde{X}_-(r, -\omega)$ from \eqref{nore2n25}, much like \eqref{phi1} before, 
in the following way:
\bg\label{bjosie}
\widetilde{X}_1(r, -\omega) ~ = ~ \alpha_1 \widetilde{X}_+(r, -\omega) ~ + ~ \alpha_2 \widetilde{X}_-(r, -\omega), \nd
as in eq (3.191) of \cite{fep} and $\alpha_i$ are, for the time being considered to be some functions of $r$ and $\omega$. The $r$ dependence of $\alpha_i$ would imply, holographically, the 
scale dependence of certain Schwinger-Keldysh parameters. The bulk viscosity may then be expressed in terms of the ratio between the individual Fourier components, 
as in eqn (3.195) of \cite{fep} which, for our case, becomes 
the ratio between ${\widetilde{X}_1'(r, -\omega)\over \widetilde{X}_1(r, -\omega)}$. Therefore using \eqref{bjosie}
the bulk viscosity $\zeta$ may be expressed as:
\bg\label{sovera}
\zeta ~& =& ~ \lim_{\omega \to 0} {{\cal G}(r) \over 2}\left[{\widetilde{X}_1'(r, -\omega)\widetilde{X}_1^\ast(r, -\omega) - \widetilde{X}_1^{\ast'}(r, -\omega) \widetilde{X}_1(r, -\omega)\over 
\vert \widetilde{X}_1(r, -\omega)\vert^2}\right]\Bigg\vert_{r_h}^{r_c} \nonumber\\
& = & ~ \lim_{\omega \to 0} {\epsilon {\cal G}(r) \over 2\omega} \left[{2(\alpha_2'\alpha_2 - \alpha_1'\alpha_2)f_0 
+ (\alpha_2^2 - \alpha_1^2) f_0'\over (\alpha_1 + \alpha_2)^2} ~ + ~ {\cal O}(\omega^2)\right]\Bigg\vert_{r_h}^{r_c}, \nd
where ${\cal G}(r)$ is derived from the background data and may be extracted from the type 
IIB action. For our case, using the coordinate system \eqref{eye} and the technique elucidated in \cite{fep}, ${\cal G}(r)$ can be expressed as:
\bg\label{snow}
{\cal G}(r) ~ = ~ r^5\sin~\theta_1~\sin~\theta_2 \left({r^2 + 9 a^2\over r^2 + 6 a^2}\right) \left[{1\over 48} - {g(r) \over 9}\right], \nd
where $a^2$ is as given in \eqref{fatur}, and $g(r)$ is the full black-hole factor including $\epsilon$ corrections. We can also compute the entropy density $s$ in terms of the parameters of 
our background. The result can be written as:
\bg\label{faihope}
s ~ = ~ {r^5\sin~\theta_1~\sin~\theta_2\over 108 T_c} \left({r^2 + 9 a^2\over r^2 + 6 a^2}\right) g'(r), \nd
where we have used the cut-off temperature $T_c$ as in \cite{fep}; and 
note the appearance of angular coordinates $\theta_1$ and $\theta_2$ as well as the resolution parameter $a^2(r)$ in a similar fashion as in ${\cal G}(r)$ above. This implies that a 
more useful thing would be to compute the {\it ratio} of the bulk viscosity \eqref{sovera} with the entropy density $s$ \eqref{faihope}. To proceed, we will then need the precise forms for 
$\alpha_1$ and $\alpha_2$ in \eqref{bjosie}. Ignoring the scale dependence of $\alpha_i$, and choosing an appropriate quadrant (see \cite{herson}) we define:
\bg\label{shysteph}
\alpha_1(\omega) ~ = ~ \alpha_0, ~~~~~~~~~ \alpha_2(\omega) ~ = ~ \alpha_0 ~e^{\omega/T_c}, \nd
with a constant $\alpha_0$ and using the same cut-off temperature $T_c$. Combining \eqref{sovera}, \eqref{snow}, \eqref{faihope} and \eqref{shysteph}, the bulk viscosity to the entropy 
ratio can be written as:
\bg\label{cremcuz}
{\zeta\over s} ~ = ~ {27 ~f_0'(r) \epsilon\over g'(r)}\left[{g(r)\over 9} - {1\over 48} \right]\Bigg\vert_{r_h}^{r_c}, \nd    
in terms of $g(r)$, whose exact value was given earlier in \eqref{bhole}, and the functional form for $f_0(r)$. Despite appearance, the ratio \eqref{cremcuz} is {\it not}
independent of $T_c$, as $T_c$ would re-emerge from the black hole factor\footnote{Recall our definition of $T_c$ in \cite{fep}: $T_c = {g'(r_h) \over 4\pi\sqrt{h(r_h)g(r_c)}}$. \label{buthib}},
but the $\omega$ dependence does cancel out 
in the final expression so that $\omega \to 0$ limit is finite. The ratio \eqref{cremcuz} is proportional to $\epsilon$, as it should be. Furthermore, to this order, we can 
replace $g(r)$ by $g_0(r)$, and ignore any corrections to $f_0(r)$ beyond zeroth order in $\epsilon$. This implies that \eqref{cremcuz} is exact to ${\cal O}(\epsilon)$.

An interesting puzzle appears at this point. Imagine we 
had chosen a different quadrant
with opposite sign for $\alpha_2$. It would naively seem that \eqref{cremcuz} cannot be finite in the $\omega \to 0$ limit as  
the bulk viscosity depends crucially on the ratio:
\bg\label{webest} {\alpha_1 - \alpha_2 \over \alpha_1 + \alpha_2}. \nd 
How can we reconcile this apparent paradox? The answer lies in the mode expansion \eqref{nore2n25}: the finiteness condition allowed us to express the Fourier modes in terms of 
$\omega^{2n-1}$. In the $\omega \to 0$ limit the ${1\over \omega}$ factor in \eqref{nore2n25}
is precisely cancelled by the ratio \eqref{webest}, as \eqref{webest} is proportional to ${\omega\over T_c}$ for 
the choice \eqref{shysteph}. However \eqref{nore2n25} is not the only choice available here.
There does exist {\it another} choice of the mode expansion that equally respects the reality condition \eqref{reality}, and can expressed as:
\bg\label{haylhaze}
\widetilde{\Gamma}_{i(\pm)} ~ = ~\left(1 \pm i\epsilon \sum_{n = 0}^\infty \widetilde{b}_{(i)n} \omega^{2n+1}\right) X_0\left(r, \vert\omega\vert\right) 
+  \epsilon X_{11}\left(r, \vert\omega\vert\right) + {\cal O}(\epsilon^2), \nd 
where $i = 1, 2, 3$ in $\widetilde{\Gamma}_i$ 
correspond to $\widetilde{T}_{\pm}$, $\widetilde{R}_{\pm}$ and $\widetilde{X}_{\pm}$ respectively; and $i = 1, 2, 3$ in $\widetilde{b}_i$ correspond to the three functions $\widetilde{p}_n(r)$, 
$\widetilde{q}_n(r)$ and $\widetilde{f}_n(r)$ respectively. As before, we will only need $\widetilde{X}_{\pm}$ and, defining $\widetilde{X}_1$ as in \eqref{bjosie} but now with 
$\alpha_2 = -\alpha_0 e^{\omega/T_c}$, the ratio of the bulk viscosity to the entropy density becomes:
\bg\label{cremcuz2}  
{\zeta\over s} ~ = ~ {108 ~T_c^2 \widetilde{f}_0'(r) \epsilon\over g'(r)}\left[{g(r)\over 9} - {1\over 48} \right]\Bigg\vert_{r_h}^{r_c}, \nd
which is finite in the $\omega \to 0$ limit as expected, and depends only on $\widetilde{f}_0$ in the series \eqref{haylhaze}. These two ratios, \eqref{cremcuz} and \eqref{cremcuz2}, are 
expected to be identical because
physical quantities cannot depend on our choice of quadrants. This turns out to be possible iff we express $T_c$ as:
\bg\label{bothkoil}
T_c ~ = ~ {1\over 2}\sqrt{f_0'(r_c)\over \widetilde{f}_0'(r_c)}, \nd
where without loss of generality we have fixed the values of $f_0$ and $\widetilde{f}_0$ at the horizon radius $r_h$. Note that \eqref{bothkoil} should be compared to the 
value of $T_c$ quoted earlier in footnote \ref{buthib}, and therfore may be used to fix the ratio $f'_0(r)/\widetilde{f}'_0(r)$ at $r_c$, the cut-off radius. Therefore with the definition of 
$T_c$, \eqref{cremcuz} or \eqref{cremcuz2} both lead to the same umabiguous value for the bulk viscosity to the entropy density ratio. 
    
\section{Discussions and Conclusions}

In this paper we performed two consistency checks and one  
computation that test the non-conformality of the model proposed in \cite{fep}. Our first consistency check is to verify the stability of the model proposed in \cite{fep}, 
elaborated later on in \cite{mia2} and 
\cite{uvcomplet}. The issue of stability arises because the UV completion in \cite{fep} requires the introduction of new degrees of freedom at a certain scale. These new degress of freedom 
appear from wrapped anti-D5 branes on two-cycle of a certain warped resolved conifold. However the presence of anti-D5 branes with wrapped D5 and D3-branes create tachyonic instabilities  
in the theory. A naive analysis demanding kappa-symmetry
along the lines of \cite{0110039} and \cite{supertube} fails because of the curvature of the cycles wrapped by the branes. Thus to restore stability we have to invoke 
non-abelian kappa-symmetry $-$ a subject that has not been developed much in the literature\footnote{We thank Eric Bergshoeff and Renata Kallosh for emphasising this.}. However despite this, we 
have been able to justify the stability of the system using certain approximate form of the non-abelian kappa-symmetry. Clearly this subject is in its infancy now, and a more detailed study 
is called for, but our preliminary analysis does shed some light on the inherent stability of the model. 

Further progress on this issue could be made by computing the higher order terms of the kappa symmetry matrix expansion in the field strength 
using the iterative procedure given in \cite{0011018v1}. An important part in our analysis was the fact that the terms at first and second order in the
 fields are either part of an expansion of a DBI-like expression or could be made to vanish on at least some subspace
 of the adjoint representation of the gauge group. For the analysis to be conclusive, a similar statement would need to be true at all orders. 
By investigating the higher order terms, or perhaps by directly analyzing the procedure that computes them, one would hope to establish
 which combinations of the non-abelian fields can appear at any given order in the expansion and check whether this is the case. 
This is an interesting problem that we relegate to future work.

Once stability is achieved, the next consistency check is the renormalization group flow from UV to IR. From the gauge theory side, this is a difficult problem as it requires the knowledge of the 
detailed behavior of the gauge theory from UV to IR. The physics is further complicated by the fact that at certain scale the theory undergoes Higgsing that converts a walking RG flow to a 
running flow that eventually leads to IR confinement, all the while remaining {\it strongly coupled}. The last requirement is an essential feature of all gauge theories that have 
gravity duals. Thus the RG flow may also be studied from the gravity dual.  
This is in principle straightforward, but in practice requires the knowledge of the background geometry and fluxes precisely. Fortunately 
the technical challenges are not insurmountable, and 
with some effort the background data may be elaborated enough leading to a complete determination of the RG flow from the gravity side. What we achieved here from the gravity dual, 
reproduces the gauge theory picture from UV to IR succinctly. Subtleties regarding strong coupling behavior, Higgsing and the smoothness of the RG flow tell us that the oft advertised 
IR Seiberg dualities etc {\it may not} be visible from the RG flow. This is of course expected, and our analysis confirms this and many other subtleties. 

Finally, we perform one computation that in some sense confirms the non-conformality in the model, namely the existence of bulk viscosity. 
We have managed to demonstrate the presence of a non-zero bulk viscosity from the gravity dual, by showing the existence of a certain imaginary piece to
the Fourier components of the 
bulk fluctuations. This imaginary piece vanishes for the conformal case, and therefore existence of this is a sure sign of 
non-conformality in the model modulo certain subtleties that we pointed out earlier. 

Interestingly, the above form \eqref{sovera} for the bulk viscosity is similar to bulk viscosity result derived from eqn (18) of \cite{gabuser}. 
Going to the $\omega \to 0$ limit, \eqref{sovera} does lead to an unambiguous value for the bulk viscosity as we show in 
\eqref{cremcuz} and \eqref{cremcuz2} by taking the ratio of the bulk viscosity with the entropy density. Our result is 
expressed in terms of certain functions $f_0(r)$ or $\widetilde{f}_0(r)$ which are related by \eqref{buthib}. These functions entail the details of the UV completion of the 
model and are therefore technically challenging to derive. At this point we
may therefore proceed in the following way.   
Using the shear viscosity
value $\eta$ presented in eqn (3.202) of \cite{fep}, the bulk to shear viscosity bound:
\bg\label{cold}
{\zeta\over \eta} ~ \ge~ 2\left({1\over 3} - c_s^2\right), \nd
with $c_s$ being the speed of sound derived from the type IIB action, may be used to find the constraint on the 
functional form for $f_0(r)$ or $\widetilde{f}_0(r)$. 
Note that the shear viscosity value in \cite{fep} uses the full UV completion, and therefore it makes sense to use \eqref{sovera} with 
contributions from the UV cap. As discussed in section \ref{coiofto}, we expect the contributions from Region 3 to be negligible, 
and the functions $f_0(r)$ and $\widetilde{f}_0(r)$ capture the contributions 
from Region 2 in \eqref{sovera}. Any additional correction that may appear from the UV cap to  ${\cal G}(r)$ given in 
\eqref{snow} will only modify $\zeta$ in \eqref{cremcuz} (or \eqref{cremcuz2}) to ${\cal O}(\epsilon^2)$, since \eqref{sovera} is already at ${\cal O}(\epsilon)$. Thus it will make sense 
to determine the constraint on $f_0(r)$ or $\widetilde{f}_0(r)$ using \eqref{cold}.
These and other details will be presented in \cite{richard}.    

\vskip.1in

\noindent {\bf Note added}: While this paper was under preparation, two papers, \cite{kirit} and \cite{mateos}, appeared in the archive that has some overlap with the contents of 
section \ref{Rogi}. It would be interesting to compare the results of \cite{kirit} and \cite{mateos} with ours. 

\vskip.15in

\centerline{\bf Acknowledgements}

\vskip.1in

\noindent We would like to thank Eric Bergshoeff, Renata Kallosh, and Mohammed Mia for helpful discussions. The work of K. D and C. G is supported in part by the Natural Sciences and Engineering 
Research Council of Canada (NSERC) grant. 
K. D would also like to thank the Simons Foundation Grant, for research support, and to Stanford University, where a substantial part of the work was done, 
for providing a stimulating research atmosphere during his sabbatical visit.


{}


\end{document}